\documentclass[12pt, useAMS, usenatbib, nomarkers, referee]{biom}

\usepackage{fancyhdr} 
\usepackage{bm}
\usepackage{amssymb}
 
\usepackage{subcaption}
\usepackage{float}
\usepackage{amsmath} 
\usepackage{dsfont}
\usepackage{thmtools} 
\usepackage{tabularx}  
\usepackage{thm-restate}
\usepackage{booktabs}
\usepackage{adjustbox}
\usepackage{xcolor}
\usepackage{hyperref}
\usepackage{url}
\usepackage{setspace}
\usepackage{multicol}
\usepackage{graphicx}
\usepackage{bbm}
\usepackage{ulem}
\graphicspath{ {./figs/} }
\usepackage{titlesec}
\titlespacing*{\subsubsection}{0pt}{0.25em}{0.25em}
\usepackage{mwe}

\usepackage{titlesec}
\usepackage{multirow}
\titlespacing*{\section}{0pt}{.25\baselineskip}{.1\baselineskip}
\titlespacing*{\subsection}{0pt}{.25\baselineskip}{.1\baselineskip}
\usepackage{graphicx} 
\allowdisplaybreaks
 \usepackage{algorithm}
\usepackage{algpseudocode}
\usepackage{bbm, dsfont}
\usepackage{soul}
\usepackage{ulem}
\usepackage[mathscr]{euscript}
\usepackage{mathrsfs}
\usepackage{enumitem}

\usepackage{enumitem}
\usepackage{authblk}
\title{FCPCA: Fuzzy clustering of high-dimensional time series based on common principal component analysis}

  \author{Ziling Ma\textsuperscript{1}, Ángel López-Oriona\textsuperscript{1}, Hernando Ombao\textsuperscript{1}, Ying Sun\textsuperscript{1}
 \thanks{King Abdullah University of Science and Technology (KAUST), Computer, Electrical and Mathematical Sciences and Engineering (CEMSE)
 Division. Thuwal 23955-6900, Saudi Arabia. 
 Correspondence:
  ziling.ma@kaust.edu.sa, angel.lopezoriona@kaust.edu.sa, hernando.ombao@kaust.edu.sa, ying.sun@kaust.edu.sa} }

\begin{document}
\pagerange{\pageref{firstpage}--\pageref{lastpage}} \pubyear{2024}

\label{firstpage}


\begin{abstract}

 Clustering multivariate time series data is a crucial task in many domains, as it enables the identification of meaningful patterns and groups in time-evolving data. Traditional approaches, such as crisp clustering, rely on the assumption that clusters are sufficiently separated with little overlap. However, real-world data often defy this assumption, exhibiting overlapping distributions or overlapping clouds of points and blurred boundaries between clusters. Fuzzy clustering offers a compelling alternative by allowing partial membership in multiple clusters, making it well-suited for these ambiguous scenarios. Despite its advantages, current fuzzy clustering methods primarily focus on univariate time series, and for multivariate cases, even datasets of moderate dimensionality become computationally prohibitive. This challenge is further exacerbated when dealing with time series of varying lengths, leaving a clear gap in addressing the complexities of modern datasets. This work introduces a novel fuzzy clustering approach based on common principal component analysis to address the aforementioned shortcomings. Our method has the advantage of efficiently handling high-dimensional multivariate time series by reducing dimensionality while preserving critical temporal features. Extensive numerical results show that our proposed clustering method outperforms several existing approaches in the literature. An interesting application involving brain signals from different drivers recorded from a simulated driving experiment illustrates the potential of the approach.  
 \\


\end{abstract}


\maketitle
\section{Introduction} \label{intro}

Clustering multivariate time series (MTS) data is a crucial task in many domains, including finance, bioscience, and environmental sciences, where understanding underlying patterns can provide valuable insights \citep{burkom2007automated, song2008tourism, ouyang2010similarity, aghabozorgi2015time}. However, the inherent complexity of MTS data poses significant challenges. Unlike univariate time series, MTS data encompass multiple interdependent variables, often exhibiting both temporal dependencies and complex cross-variable relationships, making the clustering task very challenging \citep{wang2006characteristic}. Additionally, real-world MTS datasets frequently feature variable sequence lengths and high dimensionality, further hindering the direct application of traditional clustering algorithms \citep{tucker2001variable}. These challenges are exacerbated in fuzzy clustering, where the aim is not only to group similar MTS but also to capture partial memberships, adding an additional layer of computational and methodological complexity \citep{ruspini2019fuzzy}.

Traditional crisp clustering methods, such as $K$-means, $K$-medoids, and hierarchical clustering, assign each data point to a single cluster, which makes their output very easy to interpret \citep{madhulatha2012overview,saxena2017review}. However, these methods often fail to capture nuanced relationships in complex datasets such as MTS, particularly when clusters overlap or when noise and outliers are present. Although some crisp clustering techniques allow overlapping clusters by permitting a data point to belong to multiple clusters, they still do not convey the uncertainty of data assignments \citep{jain1999data, doring2006data}. In contrast, fuzzy clustering assigns degrees of membership, allows data points to belong to multiple clusters with varying degrees of membership \citep{kruse2007fundamentals}, and effectively represents the inherent uncertainty and ambiguity in classifying data into different clusters \citep{patel2014comparative}. Moreover, the ability to assign partial memberships enables the detection of transitional states or shared characteristics between clusters, which is critical in applications, such as healthcare and finance, where these insights can have significant implications \citep{d2021trimmed,lopez2023hard}.

 In this study, we introduce a novel fuzzy clustering method tailored for moderately high-dimensional MTS, accommodating both fixed and varying lengths. In this field where research is limited, many existing approaches incur significant computational costs or fall short in managing high dimensionality \citep{maharaj1999comparison, lopez2022quantile}. In fact, in high-dimensional spaces, the very notion of similarity can become ambiguous \citep{paparrizos2017fast}. Our method introduces a framework that overcomes the limitations of traditional approaches by combining efficient dimensionality reduction with flexible handling of time series lengths. The proposed method captures both temporal dependencies and cross-variable dynamics, ensuring accurate clustering even in challenging settings.

The proposed method FCPCA leverages a fuzzy clustering framework built on common principal component analysis (CPCA) to handle MTS data. The approach begins by estimating cross-covariance matrices at different lags and constructing block matrices that combine both zero-lag and lagged dependencies. These matrices are then used to derive a common weighted (based on their corresponding membership) covariance matrix, on which singular value decomposition (SVD) is applied to extract principal components. By selecting a subset of these components that capture a high proportion of the variance, the method then projects the original MTS data onto a lower-dimensional common subspace. Membership degrees in each cluster are iteratively updated through an optimization process that minimizes the total weighted reconstruction error, with a fuzziness parameter guiding the balance between crisp and soft assignments. Furthermore, the algorithm is quite computationally efficient.

The methodology is further validated on real-world datasets, including a detailed application to an electroencephalography (EEG) simulated driving dataset. In this application, 3-second EEG segments from multiple channels are clustered to distinguish between alert and drowsy states, revealing clear groupings and transitional states that traditional hard clustering would overlook. Additional experiments on well-known datasets, such as Japanese Vowels, Basic Motions, and other benchmark datasets, demonstrate that this fuzzy clustering approach effectively captures nuanced, overlapping dynamics. This allows the algorithm to offer richer information, enabling potential integration into systems such as adaptive driver-monitoring or forecasting models.

The main contributions of this work can be summarized as follows. (1) Fuzzy clustering via lagged covariance: We extend CPCA-based clustering into a fuzzy framework, explicitly using lagged covariance blocks to better capture temporal correlations intrinsic to multivariate time series clusters. (2) Automatic fuzziness parameter selection: We propose a data-driven method to select the fuzziness parameter objectively, avoiding heuristic and subjective choices, thus improving reliability and reproducibility. (3) Membership-informed projection axes: Our method incorporates fuzzy memberships into estimating common projection axes, enhancing representativeness and inherently reducing the influence of outliers. (4) Improved performance and interoperability: FCPCA yields richer clustering insights than traditional hard clustering and consistently achieves superior accuracy compared to some existing fuzzy clustering benchmarks.

The remainder of the paper is organized as follows. Section \ref{CPCA_review} reviews related work. Section \ref{fcpca_section} details the methodology of FCPCA, including the dimensionality reduction approach, similarity measures, and fuzzy clustering framework. Section \ref{simulation} presents some numerical results. In Section \ref{application}, the proposed approach is applied to the EEG driver drowsiness dataset. Finally, Section \ref{conclusion} concludes with a discussion of the findings and potential future research directions.

\section{Hard clustering of MTS based on CPCA} \label{CPCA_review}
This section briefly reviews a clustering method for MTS based on Common Principal Component Analysis (CPCA). Although the original method (which they call Mc2PCA) was proposed by \citeauthor{li2019multivariate} (\citeyear{li2019multivariate}), we modify it by incorporating the ROBCPCA algorithm proposed by \citeauthor{ma2024robcpca} (\citeyear{ma2024robcpca}), which accounts for temporal (serial) and between time series components for MTS. These modifications enable the hard clustering method to better accommodate temporal dependencies in MTS. As suggested by the authors, we consider lag up to 2.

Consider a dataset $\bm{X} = \{\bm{X}_1, \ldots, \bm{X}_N\}$ of $N$ MTS, where $\bm{X}_i \in \mathbb{R}^{T_i \times p}$ represents the $i$th time series with $T_i$ time points and $p$ observed dimensions. Each $\bm{X}_i$ is assumed to be a realization of a $p$-dimensional second-order stationary process $\{\boldsymbol{\mathcal{X}}_t^i, t \in \mathbb{Z}\} = \{(\mathcal{X}_t^{i, 1}, \ldots, \mathcal{X}_t^{i, p})^\top, t \in \mathbb{Z}\}$, where $\mathcal{X}_t^{i, j}$ is the value of the $j$th dimension at time $t$ for the $i$th series. Second-order stationarity indicates that the mean vector $\boldsymbol{\mu}_i = E[\boldsymbol{\mathcal{X}}_t^i]$ is constant over time, and the cross-covariance matrix $\Gamma_i(l) = E[(\boldsymbol{\mathcal{X}}_t^i - \boldsymbol{\mu}_i)(\boldsymbol{\mathcal{X}}_{t-l}^i - \boldsymbol{\mu}_i)^\top]$ depends only on the lag $l \in \mathbb{Z}$ and not on $t$. The main task is to divide the $N$ MTS into $S$ non-overlapping groups.

For a given lag $l$, we estimate $\Gamma_i(l)$ using the natural sample estimate:
\[
\hat{\Gamma}_i(l) = \frac{1}{T_i - l} \sum_{t=l+1}^{T_i} (\bm{X}_{i, t} - \overline{\bm{X}}_i)(\bm{X}_{i, t-l} - \overline{\bm{X}}_i)^\top,
\]
where $\overline{\bm{X}}_i$ is the column-wise sample mean of $\bm{X}_i$.

For each lag $l \in \{1, \ldots, L\}$, a block matrix is constructed by combining lag-0 and lag-$l$ cross-covariance estimates:
\begin{equation} \label{block_cross_covariance}
    \hat{\mathbf{\Gamma}}_i(l) = 
\begin{pmatrix}
\hat{\Gamma}_i(0) & \hat{\Gamma}_i(l) \\
\hat{\Gamma}_i(l)^\top & \hat{\Gamma}_i(0)
\end{pmatrix},
\end{equation}
where $\hat{\mathbf{\Gamma}}_i(l) \in \mathbb{R}^{2p \times 2p}$. The collection of all block matrices across $N$ samples for lag $l$ is denoted as $\hat{\mathbf{\Gamma}}(l) = \{\hat{\mathbf{\Gamma}}_1(l), \ldots, \hat{\mathbf{\Gamma}}_N(l)\}$.

Next, the common matrix $\overline{\mathbf{\Sigma}}(l)$ is obtained by averaging over all $N$ samples:
\begin{equation}\label{common_matrix}
    \overline{\mathbf{\Sigma}}(l) = \frac{1}{N} \sum_{i=1}^N \hat{\mathbf{\Gamma}}_i(l).
\end{equation}

Applying SVD to $\overline{\mathbf{\Sigma}}(l)$ yields eigenvalues $\mathbf{\Lambda}^l = (\lambda_1^l, \ldots, \lambda_{2p}^l)$ and eigenvectors $\mathbf{V}^l = \{\mathbf{V}_1^l, \ldots, \mathbf{V}_{2p}^l\}$. A common projection space $\mathbf{C}(l)$ is constructed by selecting the first $k(l)$ components of $\mathbf{V}^l$, where $k(l)$ is chosen to capture at least 95\% of the total variance: 
\begin{equation}\label{num_principal_components}
    k(l) = \min \left\{ r \in \{1, \ldots, 2p\} : \frac{\sum_{j=1}^r \lambda_j^l}{\sum_{j=1}^{2p} \lambda_j^l} \geq 0.95 \right\}.
\end{equation}
The clustering procedure divides $\bm{X}$ into $S$ groups. For each cluster, the common spaces $\bm{\mathcal{C}}(l) = \{\bm{C}_1(l), \ldots, \bm{C}_S(l)\}$ are constructed. To avoid of having non-conformable arguments,  $\hat{\mathbf{X}}_i(l)$ is introduced  as
\begin{equation} \label{reconstructed_MTS_l}
    \hat{\mathbf{X}}_i(l) = (\mathbf{X}_{i,t-l}^*,\mathbf{X}_{i,t}^*),
\end{equation}
where 
$$\mathbf{X}_{i,t-l}^* = (\mathbf{X}_{i,1},\ldots, \mathbf{X}_{i,T_i-l})^\top,$$
and 
$$\mathbf{X}_{i,t}^* = (\mathbf{X}_{i,1+l},\ldots, \mathbf{X}_{i,T_i})^\top.$$
Later, each $\hat{\mathbf{X}}_i$ is projected onto the common spaces as 
\begin{equation}\nonumber \label{reconstructed_MTS}
    \quad \bm{Y}_i^s(l) = \hat{\bm{X}}_i(l) \bm{C}_s(l) \bm{C}_s(l)^\top, \quad \text{where}\:s\in\{1,\cdots, S\}.
\end{equation}
After projection, the reconstruction error $E_{is}(l)$ of projecting $i$-th MTS object $\hat{\mathbf{X}}_i(l)$ onto the $s$-th cluster at lag $l$ is computed as:
\begin{equation}   \label{reconstruction_error}
    E_{is}(l) = \lVert \hat{\bm{X}}_i(l) - \bm{Y}_i^s(l) \rVert_2.
\end{equation}
The object $\bm{X}_i$ will be grouped into the cluster where its total reconstruction error across all lags
\[
E_i = \min_{s \in \{1, \ldots, S\}} \sum_{l=1}^L E_{is}(l),
\]
is the minimum.\\
As a result, the overall error is:
\begin{equation}\label{overall_error}
    E = \sum_{i=1}^N E_i.
\end{equation}

The procedure iteratively updates the cluster partition and projection spaces until the convergence of Equation \ref{overall_error} or the maximum number of iterations.

Although CPCA-based hard clustering approaches provide valuable insight into the structure of multivariate time series data, they suffer from inherent limitations. They assign each data point to a single cluster, which can be overly restrictive, especially when the boundaries of the cluster are ambiguous or overlapped. Consequently, the fuzzy version is preferred because it allows data points to belong to multiple clusters with varying degrees of membership, providing a more flexible approach to modeling complex, overlapping structures in MTS data.

\section{Fuzzy clustering of MTS based on CPCA}\label{fcpca_section}

In this section, we extend the previously described hard clustering algorithm to a fuzzy clustering framework. We call the corresponding procedure FCPCA. Clustering is performed based on the reconstruction error of each MTS in each common space, which differs from distance-based clustering in classical fuzzy $C$-means \citep{bezdek1984fcm} or fuzzy $C$-medoids \citep{krishnapuram1999fuzzy}.

\subsection{The FCPCA clustering problem}


In this framework, we propose to perform fuzzy clustering on $\bm{X}$ by using the FCPCA algorithm. The main goal is to find the $N\times S$ matrix of fuzzy membership, $\mathbf{U} = (u_{is})$, $i=1,\ldots, N$, $s=1, \ldots, S$, and sets of common projection axes, ${\bm{\mathcal{C}}}(l) = \{{\mathbf{C}}_1(l), \ldots, {\mathbf{C}}_S(l)\}$, to minimize the total weighted reconstruction error. Formally, the clustering problem is stated as
\begin{equation}\label{fuzzyCPCA_mini}
\left\{
\begin{aligned}
    &\min_{\mathbf{U}, \tilde{\bm{\mathcal{C}}}(l)} \quad \sum_{i=1}^{N} \sum_{s=1}^{S} u_{is}^m \sum_{l=1}^{L} \| \hat{\bm{X}}_i(l) - \hat{\bm{X}}_i(l) {\mathbf{C}}_s(l) {\mathbf{C}}_s(l)^\top \|^2 \\
    &\text{subject to} \quad \sum_{s=1}^{S} u_{is} = 1 \quad \text{and} \quad u_{is} \geq 0
\end{aligned}
,\right.
\end{equation}
\noindent where $u_{is} \in [0,1]$ represents the membership degree of the $i$th MTS in the $s$th cluster, and $m > 1$ is a real number, usually referred to as the fuzziness parameter, regulating the fuzziness of the partition. For $m=1$, the crisp version of the algorithm is obtained, so the solution takes the form $u_{is}=1$ if the $i$th series pertains to cluster $c$ and $u_{is}=0$ otherwise. As the value of $m$ increases, the boundaries between clusters get softer and the resulting partition is fuzzier.

To integrate fuzzy membership degrees into the projection space, we define a common weighted covariance matrix for cluster $s$:
\begin{equation}\label{common_weighted_matrix}
   \tilde{\mathbf{\Sigma}}_s(l) = \frac{\sum_{i=1}^N u_{is}^m \hat{\mathbf{\Gamma}}_i(l)}{\sum_{i=1}^N u_{is}^m}, \quad s\in \{1,\cdots, S\},
\end{equation}
where $\hat{\mathbf{\Gamma}}_i(l)$ is the estimated block covariance matrix for the $i$th MTS object at lag $l$ given by Equation \ref{block_cross_covariance}. The set of projection axes ${\bm{\mathcal{C}}}(l) $ is then obtained by selecting principal components from each $\tilde{\mathbf{\Sigma}}_s(l)$ (e.g., retaining enough components to capture 95\% of the variance or a fixed number of components).

To solve the minimization problem above, an iterative algorithm that alternately optimizes the membership degrees and the projection axes is considered. For a given membership matrix $\mathbf{U}$, the algorithm first computes all $\tilde{\mathbf{\Sigma}}_s(l)$ and their corresponding ${\bm{\mathcal{C}}}(l) $. It then evaluates the reconstruction error for each MTS $\bm{X}_i$ under the projection space of each cluster. The membership degrees $u_{is}$ are updated by comparing reconstruction errors across clusters as follows:
\begin{equation} \label{membership_update_nonrob}
   u_{is} = \left[ \sum^S_{s^*=1} \left( \frac{\sum_{l=1}^L \| \hat{\bm{X}}_i(l) - \hat{\bm{X}}_i(l){\mathbf{C}}_{s}(l){\mathbf{C}}_{s}(l)^\top \|^2}{\sum_{l=1}^L \| \hat{\bm{X}}_i(l) - \hat{\bm{X}}_i(l){\mathbf{C}}_{s^*}(l){\mathbf{C}}_{s^*}(l)^\top \|^2} \right)^{\frac{1}{m-1}} \right]^{-1}.
\end{equation}
The total weighted reconstruction error is recalculated after updating $\mathbf{U}$. The number of principal components is set during the first iteration and remains fixed thereafter. The procedure continues until convergence (i.e., until the total reconstruction error ceases to decrease) or until a maximum number of iterations is reached.

 \subsection{The selection criteria of $m$ and $S$}
In practical applications, the fuzziness parameter $m$ is often chosen subjectively. However, the limitation of this simplistic approach can lead to downstream problems because an unsuitable choice can significantly affect the clustering outcome. Meanwhile, determining the correct number of clusters is equally important, as most clustering algorithms require it to be specified in advance, information that is rarely known in practice or requires expert knowledge of the underlying dataset \citep{kodinariya2013review}. Hence, a cluster validity index (CVI) is crucial to serve as a criterion for selecting both $m$ and the number of clusters $C$ \citep{wu2012analysis}. Here, we propose an approach that incorporates both membership degrees and dataset properties \citep{zhou2014fuzziness}, adapting the Xie--Beni index \citep{xie1991validity} as follows:
\begin{equation}\label{cpc_sep}
    \text{CVI}_{S,m} 
    \;=\;
    \frac{
      \displaystyle 
      \sum_{i=1}^{N} \sum_{s=1}^{S} u_{is}^m 
      \sum_{l=1}^{L} 
      \bigl\| \hat{\bm{X}}_i(l) \;-\; \hat{\bm{X}}_i(l)\,{\mathbf{C}}_s(l)\,{\mathbf{C}}_s(l)^\top \bigr\|^2
    }{
      \displaystyle
      N\;\min_{r \neq t} \sum_{l=1}^L 
      \bigl\| P_r \;-\; P_t \bigr\|^2
    },
\end{equation}
where
\[
P_s \;=\; {\mathbf{C}}_s(l)\,{\mathbf{C}}_s(l)^\top, 
\quad s \in \{1,\dots,S\}.
\]
Each $P_s$ is a $2p \times 2p$ matrix, regardless of the dimension of the subspace. We denote the CVI evaluated at fuzziness $m$ by $\text{CVI}_m$, so each $m$ yields a corresponding CVI value. In Equation \ref{cpc_sep}, the numerator is precisely the objective function (which we prefer to be as small as possible), while the denominator measures cluster separation (which we want to be as large as possible). Consequently, a smaller value of $\text{CVI}$ is preferred. It is also worth mentioning that all the elements used to compute the CVI value are the outputs of a given clustering solution (the total weighted reconstruction error using the specified $m$ and $S$ and the corresponding projection axes). Thus, users can perform a grid search for $m$ and $S$ with different combinations and report the results for the scenario that has the minimum CVI value.

\subsection{The FCPCA clustering procedure summary}
Algorithm~\ref{fcpca_algorithm} outlines the FCPCA clustering procedure. Before each run, the membership matrix $\mathbf{U}$ is randomly initialized (normalizing rows to ensure $\sum_s u_{is} = 1$). Because different random starts can lead to different results, we suggest repeating the entire clustering process (Steps 6--15) multiple times (3--5 in practice) and selecting the final solution that yields the smallest total weighted reconstruction error $E^*$.

\begin{algorithm}[h]
\caption{Fuzzy clustering of MTS based on CPCA: FCPCA }
\label{fcpca_algorithm}
\textbf{Input:} MTS dataset $\bm{X} = \{\bm{X}_1, \dots, \bm{X}_N\}$, number of clusters $S$, fuzziness parameter $m$, maximum number of iterations $max\_iter$, number of replicates $R$. \\
\textbf{Output:} Optimal membership matrix $\bm{U}^{*}$, projection axes ${\bm{\mathcal{C}}}^*(l)$, final total weighted reconstruction error $E^*$, the hard clustering labels vector, and the CVI value.

\begin{algorithmic}[1]
\State Normalize the MTS data: mean-center each column of $\bm{X}_i \gets$ for every unit $i$.
\State Compute the block covariance matrices for lag 1 and lag 2 using Equation \ref{block_cross_covariance}.
\State Compute $k(l)$ using Equation \ref{num_principal_components} for each lag $l$ and retain this $k(l)$ for all iterations.
\State Initialize $E^* \gets \infty$.
\State \textbf{for} $r = 1, \dots, R$ \textbf{do} \Comment{Replication loop}
    \State Initialize membership matrix $\bm{U}$ randomly.
    \State Initialize $E \gets \infty$, iteration counter $t \gets 0$.
    \State \textbf{repeat:}
    \State \quad Construct common weighted covariance matrix for each cluster using Equation \ref{common_weighted_matrix}.
    \State \quad Perform SVD on the matrices in Step 9 to obtain the common projection axes ${\bm{\mathcal{C}}}(l)$.
    \State \quad Compute reconstruction error for each MTS to each cluster using Equation \ref{reconstruction_error}.
    \State \quad Compute total weighted reconstruction error (across all clusters) $E$ using Equation \ref{fuzzyCPCA_mini}.
    \State \quad Update membership matrix $\bm{U}$ using Equation \ref{membership_update_nonrob}.
    \State \quad Increment iteration counter $iter \gets iter + 1$.
    \State \textbf{until} {$|E_{\text{prev}} - E| < \epsilon$ or $iter \geq max\_iter$}.
    \State \textbf{if} $r = 1$ \textbf{or} $E < E^*$ \textbf{then} \Comment{Select the best replication}
    \State \quad $E^* \gets E$, $\bm{U}^{*} \gets \bm{U}$, ${\bm{\mathcal{C}}}^{*}(l) \gets {\bm{\mathcal{C}}}(l)$.
\State \textbf{end for}
\end{algorithmic}
\end{algorithm}

Here, we make a short summary 
of the whole algorithm. In Step 1, each $\bm{X}_i$ is mean-centered column-wise, removing potential biases in each dimension. Steps 2--3 compute the block matrices (lagged covariance structures) and the number of principal components to retain. Steps 6--15 provide the core iterative process, updating membership degrees and re-estimating projection axes until convergence. Finally, Steps 16--18 select the best replication among $R$ runs, providing the final membership matrix $\bm{U}^*$, the projection axes ${\bm{\mathcal{C}}}^*(l)$, and the total weighted reconstruction error $E^*$. The algorithm also returns the hard clustering labels (obtained by taking the largest membership for each $\bm{X}_i$) and a CVI under a certain $m$.

Algorithm~\ref{fcpca_algorithm} is presented in a form that the user has to select $m$ beforehand, in a subjective manner. However, suppose $m$ is not specified in our implementation. In that case, we select it from the range $m \in [1.1, 2.2]$ (which the user can adjust if needed) by running the algorithm for multiple values of $m$ and returning the solution corresponding to the optimal value determined by the minimum $\text{CVI}_m$.

\section{Numerical studies} \label{simulation}

This section shows the performance of the proposed method through several simulated scenarios and some well-known MTS datasets. 

\subsection{MTS generated by VARMA process } \label{VAR_simu}

In this section, we focus on MTS data generated from Vector Autoregressive Moving Average (VARMA) processes.

\subsubsection{The VARMA model}
Ths study is designed to address the limited research on fuzzy clustering for MTS with both high dimensions and varying lengths.  VARMA models are widely used in time series analysis to capture lagged dependencies and moving average effects across multiple variables \citep{duker2025vector}. The general form of the VARMA process of order $(a, b)$ is expressed as \citep{shumway2000time}
\begin{equation}
\mathbf{X}_t = \sum_{i=1}^a \mathbf{\Phi}_i \mathbf{X}_{t-i} + \sum_{j=1}^b \mathbf{\Theta}_j \mathbf{\varepsilon}_{t-j} + \mathbf{\varepsilon}_t,
\label{eq:varma}
\end{equation}
\noindent where
\begin{itemize}
    \item $\mathbf{X}_t \in \mathbb{R}^p$ is the $p$-dimensional vector of observed time series at time $t$.
    \item $\mathbf{\Phi}_i \in \mathbb{R}^{p \times p}$ are the coefficient matrices for the autoregressive (AR) terms.
    \item $\mathbf{\Theta}_j \in \mathbb{R}^{p \times p}$ are the coefficient matrices for the moving average (MA) terms.
    \item $\mathbf{\varepsilon}_t \sim \mathcal{N}(\mathbf{0}, \mathbf{\Sigma})$ is the $p$-dimensional vector of white noise with zero mean and covariance matrix $\mathbf{\Sigma}$. Hereafter, we take $\mathbf{\Sigma}$ to be the identity matrix.
\end{itemize}

\subsubsection{Simulation setup}
\begin{figure}
    \centering
    \includegraphics[width=1\linewidth]{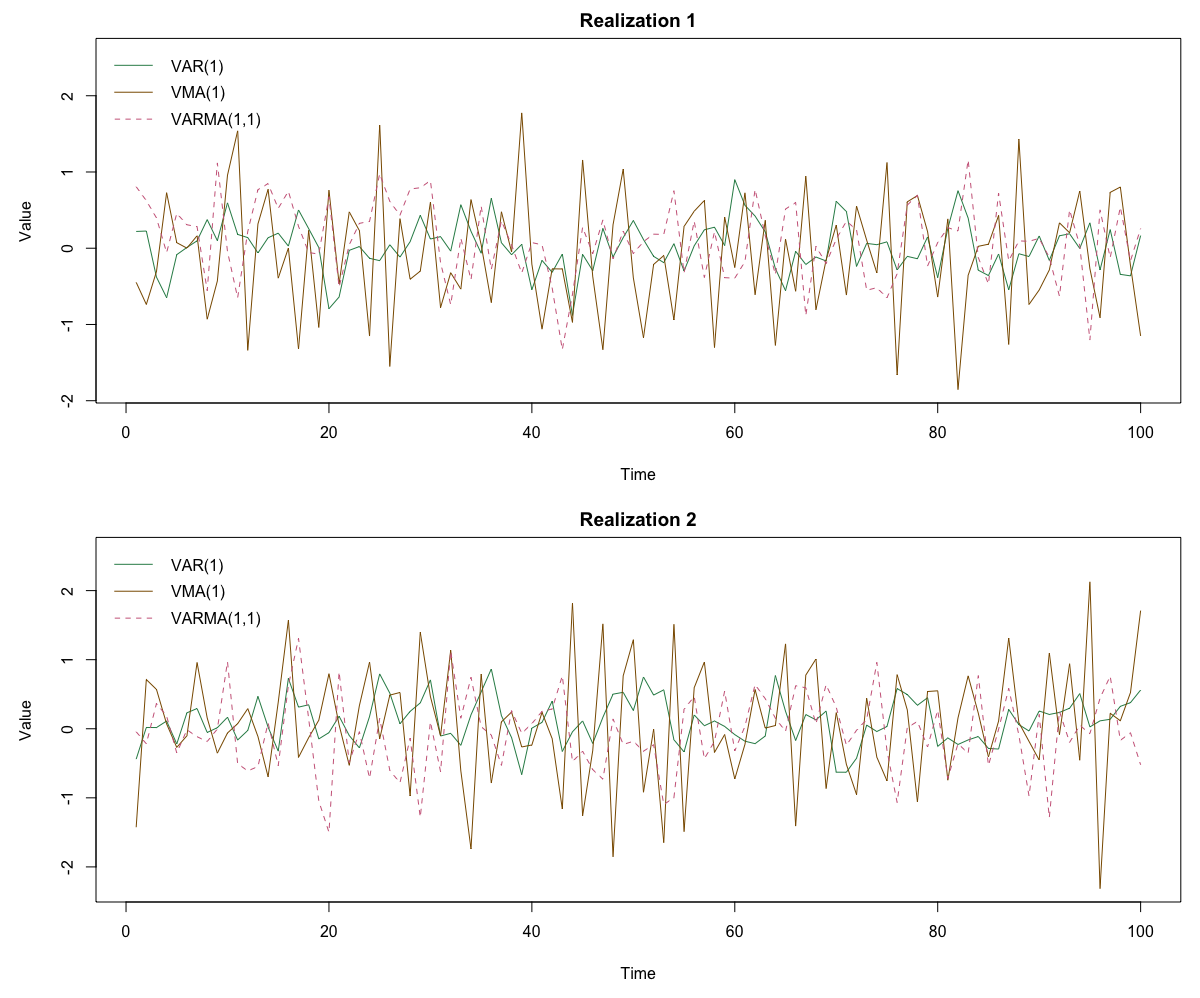}
    \caption{Example of two realizations of VARMA models. In each case, realizations of the associated VAR(1) and VMA(1) processes are also shown.}
    \label{varma_reali}
\end{figure}

We generate MTS from three distinct models: {VAR(1)}, {VMA(1)}, and {VARMA(1,1)}. Specifically, 10 series follow the VAR(1) process, 10 follow the VMA(1) process, and 2 follow the VARMA(1,1) with matrices of coefficients taken from the VAR(1) and VMA(1) models. The matrices of coefficients are generated randomly for VAR(1) and VMA(1). The {VARMA(1,1)} model, combining VAR(1) and VMA(1), can exhibit switching behavior between the two underlying processes \citep{d2012wavelets}. Two realizations are presented in Figure \ref{varma_reali}. One can see that the VARMA(1,1) series (the dashed pink line) exhibits behaviors that reflect both AR‐type persistence and MA‐type short‐lag adjustments. In other words, at times, there are occasional quick reversals or spikes similar to the pure VMA(1) series, while at other times, the series exhibits trends or drifts more like the VAR(1) process.

\noindent
We consider three scenarios with different time series lengths:
{\setlength{\parindent}{0pt}
\begin{enumerate}[leftmargin=2em]
 \item All MTS have length 200.
  \item All MTS have length 400.
   \item All MTS have lengths randomly sampled from the range 200--600.
\end{enumerate}}

In each scenario, we also vary the dimensionalities of the time series (20, 60, or 100). Thus, this setup helps to evaluate the performance of our clustering method under various lengths and dimensionalities. In all cases, the true partition is given by two groups containing the 10 elements and 1 group containing 2 elements, in accordance with the generating mechanism.

\subsubsection{Clustering and membership thresholds}

The algorithm is executed with two clusters ($S=2$). Since our method produces fuzzy memberships, we first apply a membership threshold (0.7 or 0.6) to obtain a final label for each time series. Specifically, a given time series is assigned to the cluster in which it has a membership degree above the threshold \citep{ann2010wavelet}. We expect this to occur for the well-separated VAR(1) and VMA(1) time series. In contrast, if a given time series has both membership degrees below the threshold, we label such a series as ``mixed", creating a third group. We expect this to happen for the VARMA(1,1) time series, indicating that these series do not strongly favor either cluster, reflecting the combined characteristics of VAR(1) and VMA(1) \citep{maharaj2011fuzzy}. By testing both thresholds (0.6 and 0.7), we can also assess the sensitivity of our fuzzy clustering results to the choice of the membership cutoff. 

\subsubsection{Evaluation criterion} \label{criterion_evaluate}
To quantify the agreement between the true partition and the crisp version of the clustering solution obtained as described above, we employ the Rand index (RI) \citep{rand1971objective}. The RI measures the proportion of pairwise agreements between two partitions:
\begin{equation} \label{rand_ind}
\text{RI} = \frac{\text{TP} + \text{TN}}{\text{TP} + \text{TN} + \text{FP} + \text{FN}},
\end{equation}
where \text{TP} and \text{TN} are the numbers of correctly clustered and correctly separated pairs, respectively, while \text{FP} and \text{FN} represent misclustered pairs. RI has range $[0, 1]$. Higher RI values indicate greater similarity between the experimental and true partitions. This setup allows us to evaluate our proposed clustering method under both well-separated and partially overlapping time-series dynamics. This criterion was also suggested and used by \citeauthor{li2020fuzzy} (\citeyear{li2020fuzzy}) in their paper.

\subsubsection{Competing methods}
To better understand the performance of FCPCA, we compare it with the following competing approaches:
{\setlength{\parindent}{0pt}
\begin{itemize} 
    \item \textbf{Variable-based principal component analysis (VPCA) clustering.} This method combines VPCA with a spatial-weighted matrix distance fuzzy clustering approach, capturing both value and spatial differences among features \citep{he2018unsupervised}. We include this method as a natural benchmark because it also involves dimensionality reduction. We use the \texttt{vpca\_clustering()} function from the R package \texttt{mlmts} \citep{mlmts_pack}.
    \item \textbf{Fuzzy $C$-medoids (FCMD).} This is a fuzzy variant of the well-known $K$-medoids algorithm, where each cluster is represented by a single medoid (i.e., an actual data point) rather than a centroid derived from averaging. Data points receive partial memberships in each cluster based on their distance to the medoids \citep{coppi2006fuzzy}. In our implementation, we use the function \texttt{tsclust} from the R package \texttt{dtwclust} \citep{dtwclust_pack}, and the distance is set as \texttt{dtw\_basic}.
    
\end{itemize}}
The \texttt{dtw\_basic} offers a streamlined and efficient implementation of the dynamic time warping algorithm. Focusing on the most common settings, such as symmetric step patterns and standard warping constraints—reduces computational overhead compared to more feature-rich DTW functions. Additionally, its built-in multi-threaded support via RcppParallel significantly speeds up distance calculations, making it particularly well-suited for large-scale time series clustering and classification tasks. Therefore, it is generally faster.  Precisely, we set \texttt{type = "fuzzy"}, \texttt{distance = "dtw\_basic"}, \texttt{centroid = "fcmdd"}, \texttt{fuzzy\_control(fuzziness = m))} in the \texttt{tsclust} function we use.

Note that only one replicate is used for FCPCA to obtain the RI results. This is also the case in the following paper.

\subsubsection{Results}

The clustering quality of the proposed and alternative approaches is assessed as follows. For each simulation scenario and time series dimension, we randomly generate three distinct sets of coefficient matrices that satisfy the stationarity condition for the VAR(1) and VMA(1) processes. For each set, we simulate MTS processes 100 times, compute the Rand Index (RI) for each method at each simulation, and then take the per-method average across all 300 runs (3 sets × 100 runs). For the three methods, this is done for several values of the fuzziness parameter. 

The clustering performance of the three approaches is shown in Tables \ref{tab:FCPCA_comparison_7} and \ref{tab:FCPCA_comparison_6} for thresholds 0.7 and 0.6, respectively. The headers of each column refer to the number of dimensions of the time series. Since VPCA is unable to handle MTS with various lengths, we do not consider this approach in the third scenario. The best mean performance in each setting is marked in bold.

\begin{table}[htbp]
\centering
\small
\setlength{\tabcolsep}{10pt}
\renewcommand{\arraystretch}{1.1}
\caption{Performance (mean RI) of FCPCA, VPCA, and FCMD for different lengths, fuzziness parameters ($m$), and dimensions, with membership threshold $0.7$.}
\begin{tabular}{cc|ccc|ccc|ccc}
\toprule
\multirow{2}{*}{Length} & \multirow{2}{*}{Fuzziness ($m$)} 
& \multicolumn{3}{c|}{FCPCA} 
& \multicolumn{3}{c|}{VPCA} 
& \multicolumn{3}{c}{FCMD} \\
\cmidrule(lr){3-5}\cmidrule(lr){6-8}\cmidrule(lr){9-11}
 &  & 20 & 60 & 100 & 20 & 60 & 100 & 20 & 60 & 100 \\
\midrule
\multirow{6}{*}{200}
 & 1.2 
    & \textbf{0.91} & \textbf{0.91} & \textbf{0.91}   
   & 0.55 & 0.55 & 0.55   
   & 0.39 & 0.32 & 0.32   
   \\
 & 1.4
    & \textbf{0.91} & \textbf{0.91} & \textbf{0.91}  
   & 0.55 & 0.55 & 0.55   
   & 0.23 & 0.32 & 0.39   
   \\
 & 1.6
    & \textbf{0.91} & \textbf{0.91} & \textbf{0.91}   
   & 0.55 & 0.55 & 0.55   
   & 0.23 & 0.50 & 0.32   
   \\
 & 1.8
  & \textbf{0.92} & \textbf{1.00} & \textbf{1.00}  
   & 0.55 & 0.55 & 0.55   
   & 0.30 & 0.99 & 0.95   
   \\
 & 2.0
   & \textbf{0.90} & \textbf{0.53} & 0.38   
   & 0.37 & 0.27 & \textbf{0.55}   
   & 0.23 &\textbf{0.32} & 0.39   
   \\
 & 2.2
   & 0.10 & 0.10 & 0.10   
   & 0.10 & 0.10 & 0.27   
   & \textbf{0.14} & 0.23 & \textbf{0.50}   
   \\
\midrule
\multirow{6}{*}{400}
 & 1.2
     & \textbf{0.91} & \textbf{0.91} & \textbf{0.91}
    & 0.55 & 0.55 & 0.55   
   & 0.39 & 0.30 & 0.50
   \\
 & 1.4
     & \textbf{0.91} & \textbf{0.91} & \textbf{0.91}
    & 0.55 & 0.55 & 0.55   
   & 0.27 & 0.37 & 0.21
   \\
 & 1.6
     & \textbf{0.91} & \textbf{0.91} & \textbf{0.91}
    & 0.55 & 0.55 & 0.55   
   & 0.32 & 0.50 & 0.32
   \\
 & 1.8
   & \textbf{0.91} & \textbf{1.00} & \textbf{1.00}
    & 0.55 & 0.55 & 0.55   
   & 0.32 & 0.37 & 0.32
   \\
 & 2.0
   & 0.46 & 0.27 & 0.36
   & \textbf{0.53} & \textbf{0.52} & \textbf{0.55}
   & 0.32 & 0.41 & 0.30
   \\
 & 2.2
   & 0.13 & 0.10 & 0.10
   & 0.18 & 0.10 & 0.10
   & \textbf{0.50} & \textbf{0.39} & \textbf{0.37}
   \\
\midrule
\multirow{6}{*}{200--600}
 & 1.2
   & \textbf{0.91} & \textbf{0.91} & \textbf{0.91}
   & NA & NA & NA
   & 0.35 & 0.26 & 0.26
   \\
 & 1.4
    & \textbf{0.91} & \textbf{0.91} & \textbf{0.91}
     & NA & NA & NA
   & 0.26 & 0.21 & 0.17
   \\
 & 1.6
   & \textbf{0.91} & \textbf{0.91} & \textbf{0.91}
    & NA & NA & NA
   & 0.20 & 0.17 & 0.17
   \\
 & 1.8
  & \textbf{1.00} & \textbf{0.99} & \textbf{1.00}
  & NA & NA & NA
   & 0.25 & 0.28 & 0.25
   \\
 & 2.0
   & \textbf{0.46} & \textbf{0.90} & \textbf{0.59}
    & NA & NA & NA
   & 0.24 & 0.22 & 0.21
   \\
 & 2.2
   & 0.14 & 0.10 & 0.10
     & NA & NA & NA
   & \textbf{0.33} &\textbf{0.26} & \textbf{0.17}
   \\
\bottomrule
\end{tabular}
\label{tab:FCPCA_comparison_7}
\end{table}

\begin{table}[htbp]
\centering
\small
\setlength{\tabcolsep}{10pt}
\renewcommand{\arraystretch}{1.1}
\caption{Performance (mean RI) of FCPCA, VPCA, and FCMD for different lengths, fuzziness parameters ($m$), and dimensions, with membership threshold $0.6$.}
\begin{tabular}{cc|ccc|ccc|ccc}
\toprule
\multirow{2}{*}{Length} & \multirow{2}{*}{Fuzziness ($m$)} 
& \multicolumn{3}{c|}{FCPCA} 
& \multicolumn{3}{c|}{VPCA} 
& \multicolumn{3}{c}{FCMD} \\
\cmidrule(lr){3-5}\cmidrule(lr){6-8}\cmidrule(lr){9-11}
 &  & 20 & 60 & 100 & 20 & 60 & 100 & 20 & 60 & 100 \\
\midrule
\multirow{6}{*}{200}
 & 1.2 
   & \textbf{0.91} & \textbf{0.91} & \textbf{0.91}   
   & 0.49 & 0.52 & 0.46   
   & 0.35 & 0.28 & 0.28   
   \\
 & 1.4
   & \textbf{0.91} & \textbf{0.91} & \textbf{0.91}   
   & 0.50 & 0.55 & 0.55   
   & 0.23 & 0.28 & 0.35   
   \\
 & 1.6
   & \textbf{0.91} & \textbf{0.91} & \textbf{0.91}    
   & 0.55 & 0.55 & 0.55   
   & 0.22 & 0.43 & 0.28   
   \\
 & 1.8
   & \textbf{0.91} & \textbf{0.91} & \textbf{0.91}    
   & 0.55 & 0.55 & 0.55   
   & 0.40 & 0.43 & 0.27   
   \\
 & 2.0
  & \textbf{0.91} & \textbf{0.98} & \textbf{0.91}   
   & 0.55 & 0.48 & 0.55   
   & 0.36 & 0.20 & 0.21   
   \\
 & 2.2
   & \textbf{0.27} & 0.10 & 0.10   
   & 0.18 & \textbf{0.27} & {0.27} 
   & 0.14 & 0.21 & \textbf{0.43}   
   \\
\midrule
\multirow{6}{*}{400}
 & 1.2
   & \textbf{0.91} & \textbf{0.91} & \textbf{0.91}
   & 0.52 & 0.47 & 0.46
   & 0.35 & 0.27 & 0.43
   \\
 & 1.4
 & \textbf{0.91} & \textbf{0.91} & \textbf{0.91}
    & 0.55 & 0.55 & 0.55
   & 0.21 & 0.34 & 0.20
   \\
 & 1.6
   & \textbf{0.91} & \textbf{0.91} & \textbf{0.91}
    & 0.55 & 0.55 & 0.55
   & 0.28 & 0.34 & 0.28
   \\
 & 1.8
    & \textbf{0.91} & \textbf{0.91} & \textbf{0.91}
    & 0.55 & 0.55 & 0.55
   & 0.28 & 0.34 & 0.27
   \\
 & 2.0
   & 0.42 & \textbf{0.64} & \textbf{0.64}
   & \textbf{0.55} & 0.55 & 0.55
   & 0.28 & 0.35 & 0.27
   \\
 & 2.2
   & 0.45 & 0.45 & 0.38
   & \textbf{0.66} & \textbf{0.65} & \textbf{0.63}
   & 0.43 & 0.35 & 0.34
   \\
\midrule
\multirow{6}{*}{200--600}
 & 1.2
   & \textbf{0.91} & \textbf{0.91} & \textbf{0.91}
     & NA & NA & NA
   & 0.35 & 0.25 & 0.25
   \\
 & 1.4
    & \textbf{0.91} & \textbf{0.91} & \textbf{0.91}
     & NA & NA & NA
   & 0.25 & 0.21 & 0.17
   \\
 & 1.6
    & \textbf{0.95} & \textbf{0.91} & \textbf{0.91}
     & NA & NA & NA
   & 0.20 & 0.17 & 0.17
   \\
 & 1.8
   & \textbf{1.00} & \textbf{0.91} & \textbf{0.91}
     & NA & NA & NA
   & 0.25 & 0.28 & 0.24
   \\
 & 2.0
  & \textbf{0.82} & \textbf{0.91} & \textbf{0.99}
     & NA & NA & NA
   & 0.24 & 0.21 & 0.21
   \\
 & 2.2
   &\textbf{0.32} & 0.10 & 0.10
     & NA & NA & NA
   &  0.31 & \textbf{0.25} & \textbf{0.17}
   \\
\bottomrule
\end{tabular}
\label{tab:FCPCA_comparison_6}
\end{table}

When the fuzziness parameter is set to a high value (e.g., $m=2.2$), the accuracy of FCPCA compared to FCMD and VPCA decreases, as overly diffuse memberships weaken cluster separations. However, at more moderate fuzziness levels (around 1.2–1.9), FCPCA clearly outperforms the competing methods, regardless of the lengths and dimensionalities of the MTS data. We observe that the proposed method frequently achieves a RI of 0.91. A detailed examination of the obtained partitions reveals that, in such instances, the method fails to detect the two fuzzy series at the corresponding threshold, although it works perfectly for the well-separated groups. Therefore, an additional measure must be provided to quantify the success rate of detecting the fuzzy series. Specifically, we consider the average number of VARMA(1, 1) series that are correctly recognized as fuzzy out of the two. Table \ref{Num_detect_fuzzy} presents the results for FCPCA using the optimal value of $m$ (see Section \ref{fps}). The results for the competing approaches are not shown, as these methods nearly always achieve a success rate of 0. As a reference, the most selected $m$ for this simulation across all scenarios is mostly 1.7 or 1.8. We see that FCPCA can detect the fuzzy series in most cases, although at a threshold of 0.6, the results appear less ideal. This highlights the influence of the chosen threshold on detection performance.

\begin{table}[htbp]
\centering
\caption{Average number of fuzzy series detected out of two using the optimal $m$ 
         for method FCPCA with thresholds 0.7 and 0.6.}
\label{Num_detect_fuzzy}
\begin{tabular}{c c c c c}
\toprule
 & & \multicolumn{3}{c}{Dim}\\
\cmidrule(lr){3-5}
{Threshold} & {Length} & {20} & {60} & {100} \\
\midrule
\multirow{3}{*}{0.7} 
  & 200       & 0.24 & 2.00 & 2.00 \\
  & 400       & 0.00 & 2.00 & 2.00 \\
  & 200--600  & 2.00 & 1.78 & 2.00 \\
\midrule
\multirow{3}{*}{0.6} 
  & 200       & 0.00 & 1.55 & 0.00 \\ 
  & 400       & 0.00 & 0.00 & 0.00 \\   
  & 200--600  & 2.00 & 0.00 & 1.77  \\   
\bottomrule
\end{tabular}
\end{table}

\subsection{EEG data simulated by AR(2) mixture model}
In this section, we simulate EEG data using AR(2) mixing processes, chosen for their ability to generate oscillatory signals with well-defined peaks and spectral bandwidths. EEG, a widely used non-invasive measure of brain activity, captures essential neural dynamics \citep{sanei2013eeg}. By replicating these rhythms, AR(2) mixing processes provide realistic models for studying cognitive mechanisms and neurological disorders. Specifically, MTS are generated by AR(2) mixing processes that represent five distinct EEG frequency bands (delta, theta, alpha, beta, and gamma). The AR(2) coefficients for each band are designed to capture its characteristic oscillatory properties. A more detailed explanation of the simulation mechanism for EEG data using the AR(2) mixture model can be found in the paper by \citeauthor{ombao} (\citeyear{ombao}).

\begin{figure}[htbp]
    \centering
    \includegraphics[width=1.0\linewidth]{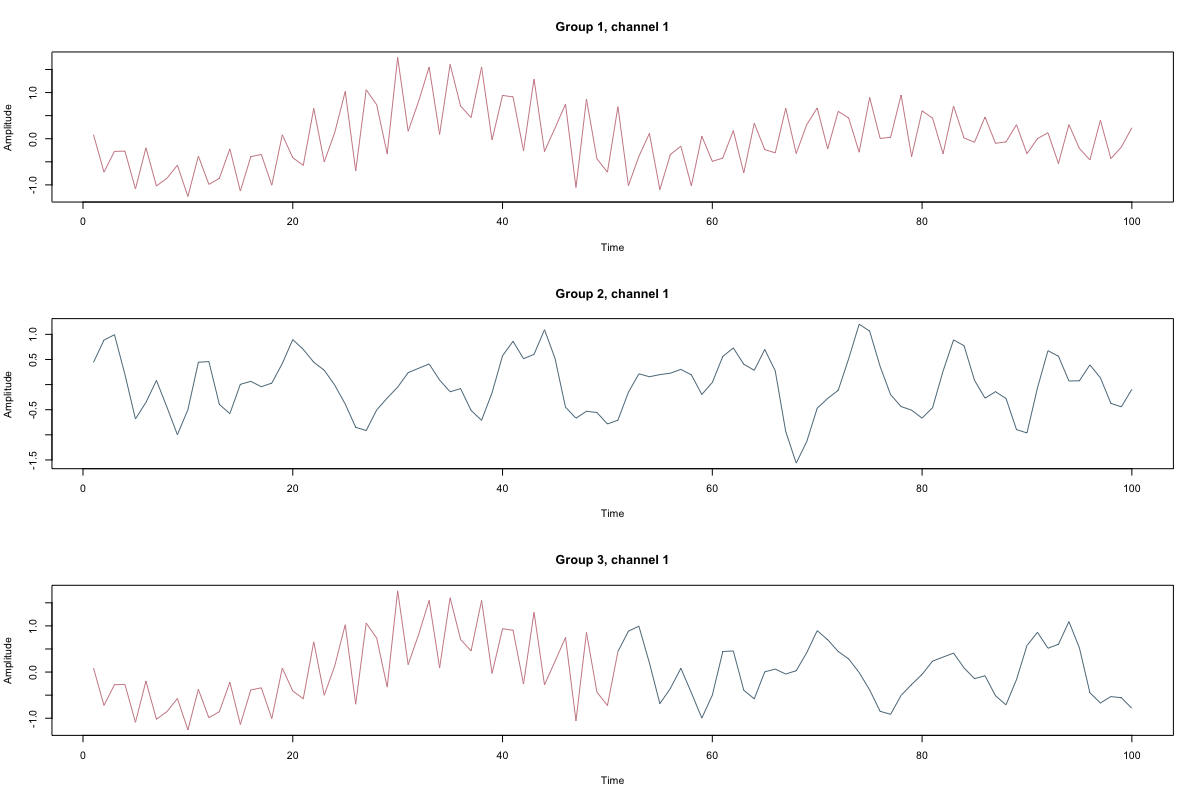}
    \caption{The simulated EEG data example.}
    \label{EEG_example}
\end{figure}

In our simulations, we generate MTS with two lengths (256 and 512), each with three different channel counts (32, 64, and 128). These lengths and channel configurations align with typical EEG sampling rates (e.g., 256 or 512 Hz) and commonly used EEG channel setups. We consider two well-separated groups. Group 1 is dominated by delta and gamma frequencies, and group 2 is dominated by theta, alpha, and beta frequencies. We then form a third, “fuzzy” group by combining the first half of the MTS from group 1 with the first half of the MTS from group 2. An example is given in Figure \ref{EEG_example}. Thus, group 3 combines half group 1 and half group 2. Each group contains 10 MTS.
We consider a threshold of 0.7 and conduct 100 simulation replicates. The results are presented in Table~\ref{RI_eeg}, where we report both the mean RI and the mean number of fuzzy series successfully detected (out of ten) using the optimal fuzziness parameter. Again, only the FCPCA results are shown, as the competing methods fail to produce meaningful outcomes regardless of the value of the fuzziness parameter. The most selected $m$ here across all scenarios is 1.2. We can see that the clustering performance of FCPCA is quite consistent, considering the different lengths and dimensions of the MTS. Our experimental results indicate that selecting a higher threshold (e.g., 0.8) can further improve clustering accuracy in this scenario.

{\singlespacing
\begin{table}[htbp]
    \centering
    \caption{The mean RI and number of fuzzy series successfully detected for FCPCA in the simulated EEG data with threshold 0.7.}
     \begin{tabular}{c|ccc|ccc}  
     \hline
    Length & \multicolumn{3}{c|}{256} & \multicolumn{3}{c}{512} \\
     \hline
  Channels  &  32 & 64 & 128 & 32 & 64 & 128 \\
  \hline
    RI   &  0.84 & 0.86 & 0.82 & 0.89 & 0.85 & 0.87 \\
    \hline 
    Number of fuzzy series detected & 5.10 & 5.24 & 5.78 &  6.70 &  5.44 & 6.04\\
\hline
\end{tabular}
    \label{RI_eeg}
\end{table}
}

\subsection{Evaluating FCPCA in a hard/crisp clustering context}
Since FCPCA is a fuzzy generalization of the CPCA-based method, we aim to assess the benefits of introducing fuzziness. Mc2PCA and ROBCPCA employ a hard clustering approach, whereas FCPCA assigns degrees of membership to each cluster. To facilitate a direct comparison with the hard clustering baseline, we use a relatively low fuzziness parameter, namely $m=1.1$, so that the memberships remain nearly crisp. In addition, we convert the resulting fuzzy partition into a hard one using the row-wise maximum rule, assigning each series to the cluster with the highest membership. Table \ref{data_descrip} provides a brief overview of the real datasets used in this study, which are commonly considered to evaluate the performance of various clustering algorithms (all of them have associated true partitions). The RI (Equation \ref{rand_ind}) is applied again for evaluation. The results presented in Table \ref{real_data_hard} were obtained by averaging the performance of each method over 100 replications for each dataset.

Interestingly, Table \ref{real_data_hard} shows that FCPCA achieves a higher clustering accuracy than the original hard clustering methods in several real-world datasets when a low value for the fuzziness parameter is considered. This improvement could suggest that even a small degree of fuzziness can help capture the underlying uncertainties in the data, leading to more robust and accurate cluster assignments.

\begin{table}[htbp]
     \caption{Datasets brief description.}
    \label{data_descrip}
    \centering
\small 
\setlength{\tabcolsep}{18pt} 
\renewcommand{\arraystretch}{1.1} 
    \begin{tabular}{c|c c c c  }
    \hline
      Dataset& Size &  Length  & Dimension  & Number of classes     \\ 
      \hline
         Articulary word recognition   & 300  & 144 &  9 & 25  \\
         \hline
      Atrial fibrillation   & 15  & 600 &  2 & 3  \\
      \hline
       Basic motions     &  40 & 100  & 6 & 4\\
      \hline
      Japanese vowels   &370 & 29& 12 & 9   \\
       \hline
       NATOPS & 180 & 51 & 24 & 6 \\
       \hline 
      Racket sports  & 152 & 30 & 6  &  4   \\
      \hline
      
    \end{tabular}

\end{table}

\begin{table}[htbp]
     \caption{The RI on different datasets with standard deviation in the bracket}
    \label{real_data_hard}
    \centering
\small 
\setlength{\tabcolsep}{12pt} 
\renewcommand{\arraystretch}{1.1} 
    \begin{tabular}{c|c c c c c}
    \hline
      Dataset& Mc2PCA & ROBCPCA  &FCPCA  & VPCA    & FCMD \\ 
    \hline 
       Articularly word recognition   & 0.17 (0.02) &  0.21 (0.02) &  0.50 {(0.01)} & \textbf{0.77} (0.04) & 0.66 (0.05) \\
       \hline
     AtrialFibrillation  & 0.38 (0.11) & \textbf{0.49} (0.06) & 0.47 {(0.00)}   & 0.44 (0.04) & 0.47 (0.06) \\
      \hline
        Basic motions     & 0.38 (0.04) & 0.46 (0.07)  & \textbf{0.79} (0.04)  & 0.46 (0.05) & 0.50 (0.14)\\
      \hline 
      Japanese Vowels   & 0.36 (0.07) & 0.57 (0.06) & 0.75 {(0.04)}  & \textbf{0.79} (0.06)  & 0.23 (0.21) \\
      \hline
       NATOPS  & 0.44 (0.09) & 0.47 (0.08) & \textbf{0.49}  {(0.02)}  & 0.45 (0.17)  & 0.46 {(0.01)} \\
      \hline
      Racket sports  & 0.31 (0.06) & 0.33 (0.03) & \textbf{0.47} (0.03)  &  0.44 (0.09)  & 0.43 (0.04)\\
      \hline
      
    \end{tabular}

\end{table}

\section{Application of FCPCA to EEG drowsiness data} \label{application}
Driver drowsiness is one of the leading causes of road fatalities \citep{brown2013identifying}, and EEG analysis provides a direct window into the neurophysiological activities of the brain
that signal the onset of fatigue. This analysis is critical because, despite the susceptibility of EEG to inter-subject variability (including mental and physical drifts), it remains one of the most informative signals for detecting drowsiness. Ultimately, analyzing EEG drowsiness data can deepen our understanding of the neurocognitive transitions from alertness to drowsiness, which is essential for designing interventions that can enhance driver safety and prevent transportation hazards.

 The dataset we use for the study is available on the website \url{https://figshare.com/articles/dataset/EEG_driver_drowsiness_dataset/14273687?file=30707285}. It contains 2022 EEG samples of size $384 \times 30$ from 11 subjects with labels of alert and drowsy. Each sample is a 3-second EEG data with 128Hz from 30 EEG channels. In this dataset, participants operated a simulated vehicle, attempting to maintain their position in the center of a lane. Lane-departure events were randomly introduced to mimic real-world conditions, such as slight road curvatures or small obstacles, causing the car to drift left or right. Each lane-departure event (a “trial”) consisted of a baseline period, the moment the car started drifting (deviation onset), the steering response of participant (response onset), and the time when the car returned to the center (response offset). Because the task was monotonous, participants often became drowsy. For each trial, a 3-second EEG segment was extracted immediately before the deviation onset, capturing the brain state leading up to the drift. A detailed description of this dataset is provided in \cite{cui2022compact}.

As mentioned by \citeauthor{zhang2023subject} (\citeyear{zhang2023subject}), since EEG signals have high variability and instability, data can differ substantially across subjects. Thus, we decided to split the whole dataset into 11 groups, each containing only the EEG signals from one subject. Later, our FCPCA algorithm is applied to each individual set of signals considering two groups, $S=2$. 

\begin{figure}
    \centering
    \includegraphics[width=0.8\linewidth]{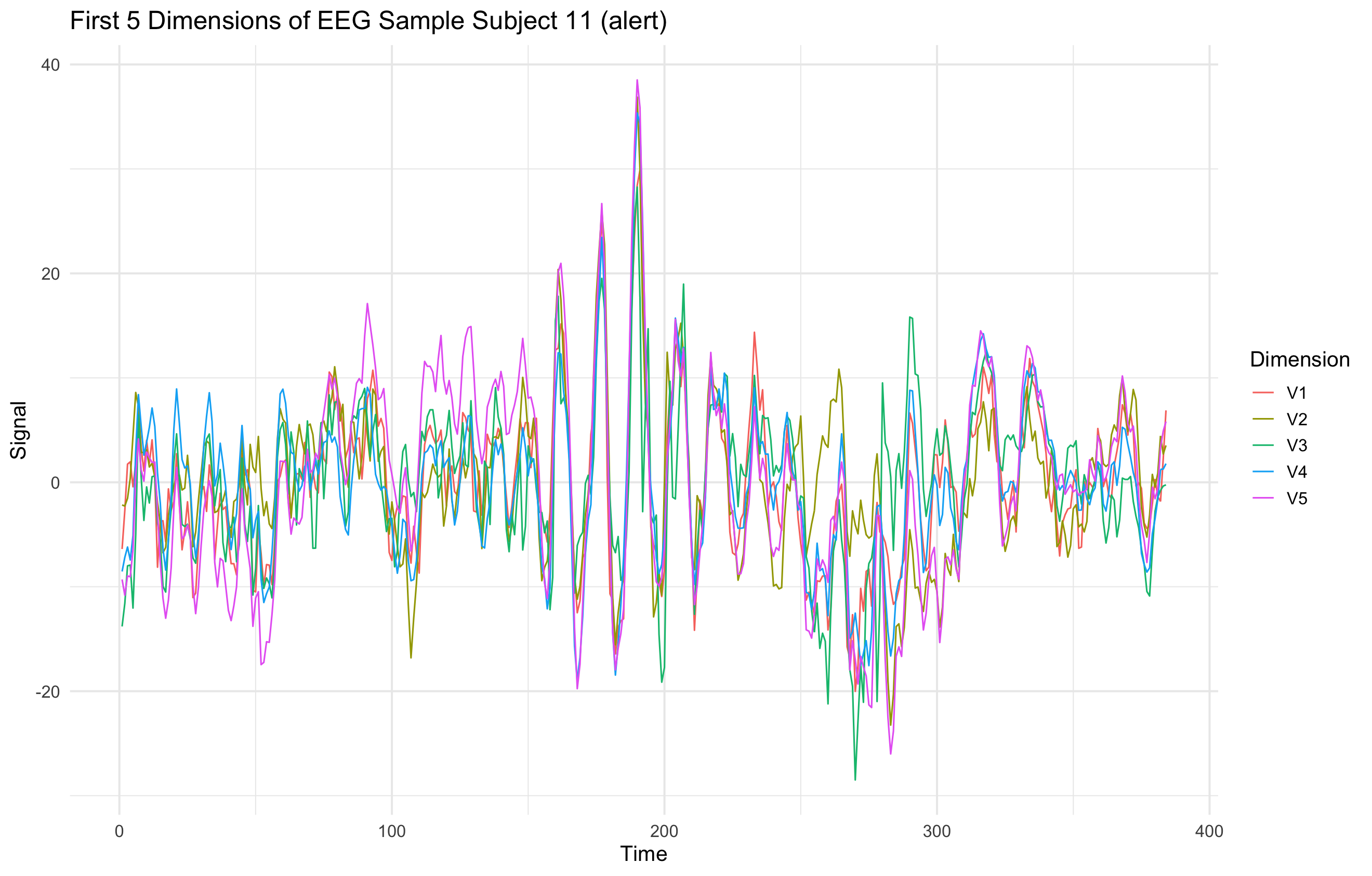}
    \caption{The example EEG signals from the first five channels of subject 11 in the alert state.}
    \label{alert}
\end{figure}

\begin{figure}
    \centering
    \includegraphics[width=0.8\linewidth]{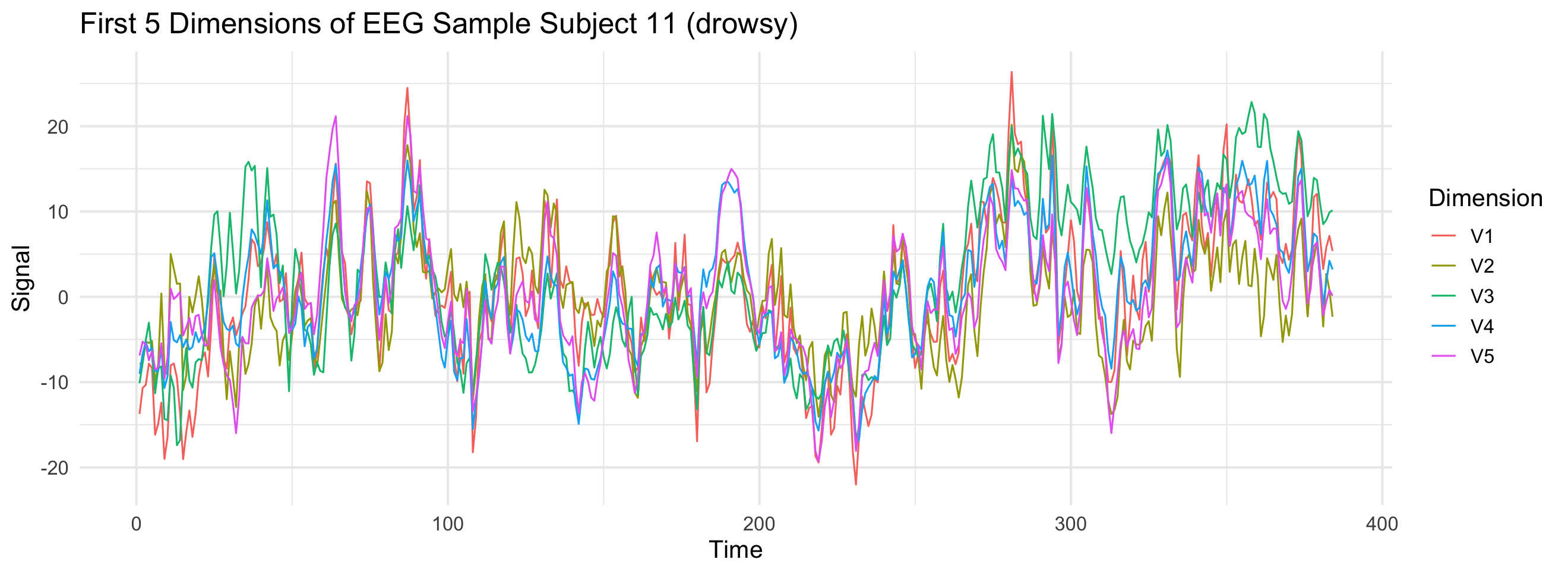}
    \caption{The example EEG signals from the first five channels of subject 11 in the drowsy state.}
    \label{drowsy}
\end{figure}

\begin{figure}[htbp]
    \centering
    \includegraphics[width=1.0\linewidth]{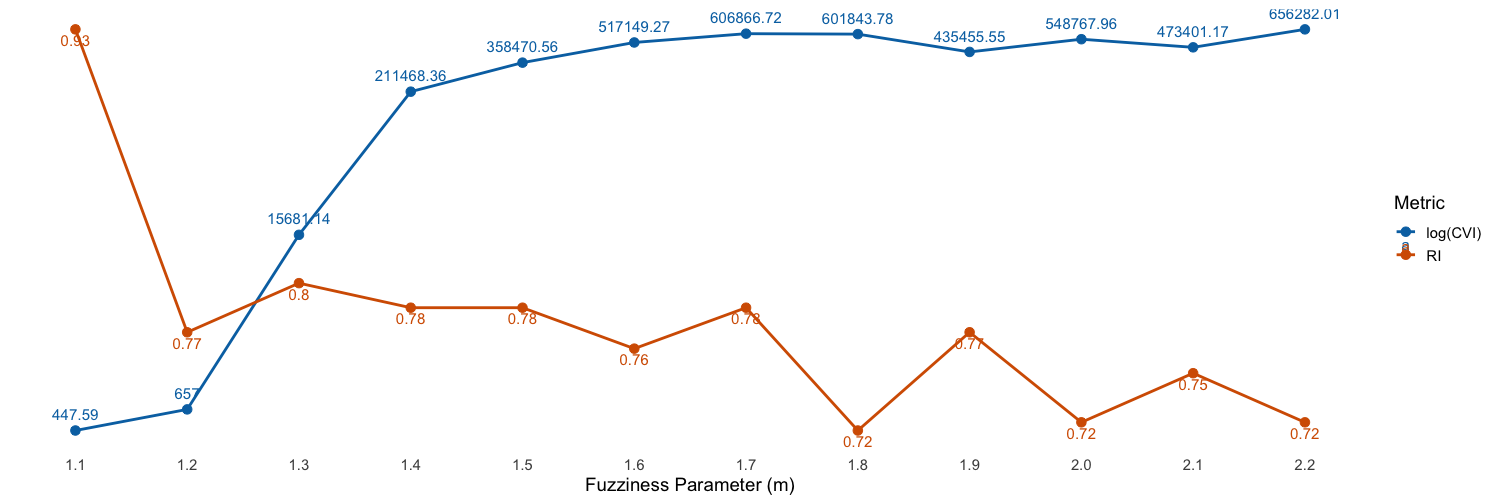}
    \caption{The selection of $m$ and comparison to the CVIs and RIs.}
    \label{m_choice}
\end{figure}

Here, we first present the results for subject number 11 from the original dataset. Example EEG signals of subject 11 under alter and drowsy state are shown in Figure \ref{alert} and \ref{drowsy}. Only the first five channels are displayed for clarity. In particular, 226 samples were collected for this subject. Using Equation \ref{cpc_sep}, we select the optimal value of \( m = 1.1 \), as it yields the minimum CVI. Figure \ref{m_choice} shows the log-transformed CVI values (employed to normalize the large CVI values) along with the corresponding RIs for various selections of \( m \). The original CVIs and RIs are marked in the figure. Notably, the optimal \( m \) also produces the highest RI, achieving a value of 0.93. However, the RI exhibits fluctuations as \( m \) increases, and the membership matrix \(\mathbf{U}\) remains approximately balanced, with membership degrees near 0.5 for both clusters. In certain instances, the RI appears to be influenced by random assignment due to identical membership degrees. 

\begin{figure}[htbp]
    \centering
    \includegraphics[width=1\linewidth]{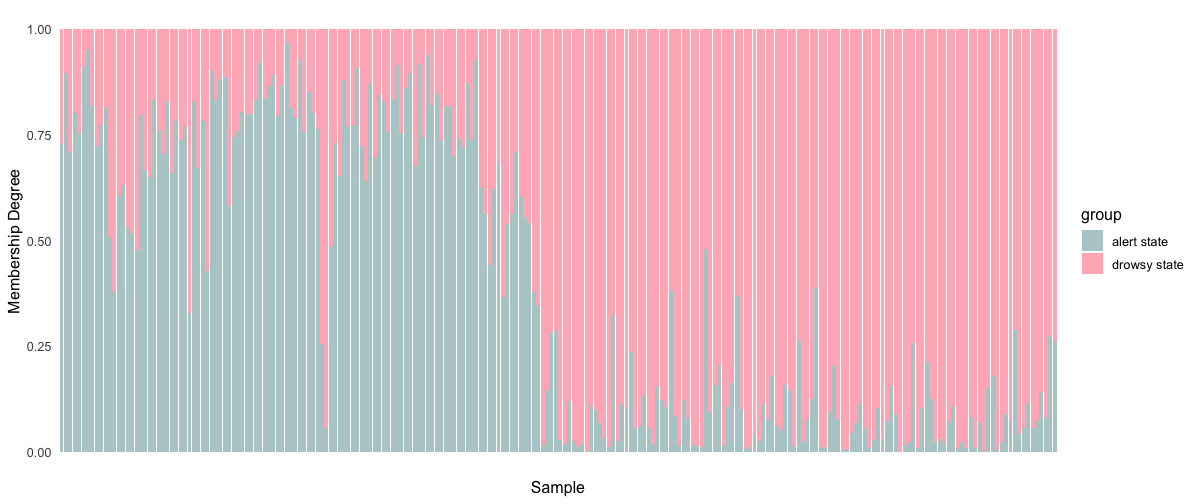}
    \caption{Membership matrix of the samples of subject 11.}
    \label{EEG_mem_matrix}
\end{figure}

The resulting membership degree structure is shown in Figure \ref{EEG_mem_matrix}. As one can see, some samples exhibit nearly crisp membership in either the drowsy or the alert cluster. For instance, a high drowsy state membership might occur when a driver travels along a monotonous route with little traffic, where boredom can set in and induce drowsiness. Conversely, a high alert state membership may appear when the driver encounters heavier traffic, needs to stay vigilant to avoid collisions, or notices engaging stimuli on road signs or in music.

More importantly, our fuzzy clustering algorithm reveals that certain samples exhibit substantial memberships in both clusters, indicating a transitional or mixed state between drowsiness and alertness. This phenomenon cannot be captured by hard clustering, which forces each sample into a single cluster. However, it is realistic to assume that drivers experience a continuum of alertness, transitioning gradually from an alert state to a drowsy state (and vice versa) over the course of a long trip. For instance, a driver who has been on the road for an extended period may start feeling fatigued but then becomes partially re-alerted upon noticing a lane drift or other stimuli, requiring additional time to regain full attentiveness.

\begin{figure}[htbp]
    \centering
    \includegraphics[width=1\linewidth]{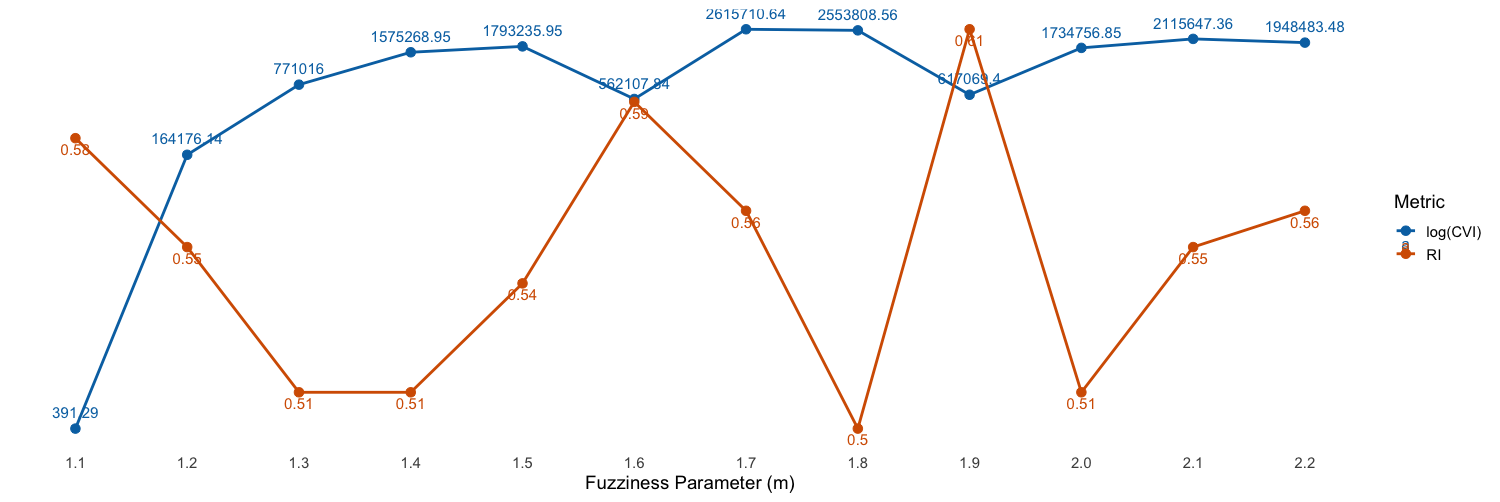}
    \caption{The selection of $m$ and comparison to the CVIs and RIs of subject 7}
    \label{subject7}
\end{figure}

\begin{figure}[htbp]
    \centering
    \includegraphics[width=1.0\linewidth]{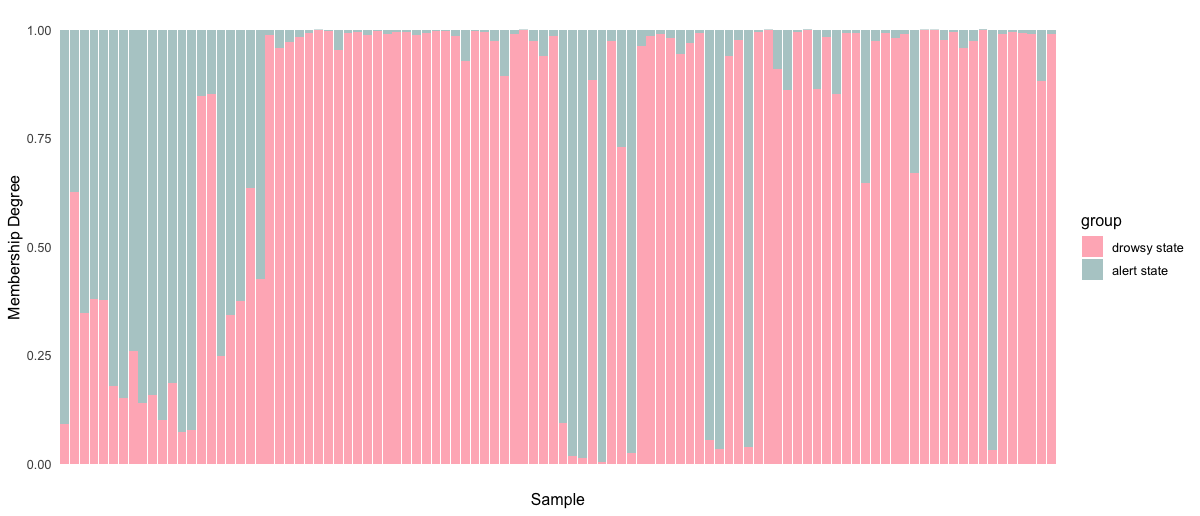}
    \caption{Membership matrix of the samples of subject 7.}
    \label{mem_7}
\end{figure}

Conversely, for subject 7, the proposed method exhibits a low clustering accuracy (0.58), with most samples classified as “drowsy.” In total, 102 samples are collected for this subject. Figures \ref{subject7} and \ref{mem_7} show the selection of $m$ and the corresponding membership matrix, respectively. 
Several factors may contribute to this outcome. First, EEG signals of this participant may lack clear-cut distinctions between drowsy and alert states, causing the algorithm to group most observations into a single cluster. Second, the data could be noisy or ambiguous, especially if the subject frequently occupies borderline states that do not neatly fit a strict binary label. Finally, the ground truth labels in this dataset are assigned using strict threshold-based rules on reaction times \citeauthor{wei2018toward} (\citeyear{wei2018toward}), which can mislabel borderline or transitional states by forcing them into one of two categories. Such rules may also misalign physiological responses with the assigned labels, particularly if reaction times do not accurately reflect true drowsiness. Consequently, the so-called “true” cluster may not fully capture the actual cognitive state.

 In short, by accommodating partial memberships, our fuzzy clustering approach provides a more nuanced view of driver states, capturing these transitional phases that hard clustering would overlook. This capability is especially valuable for designing adaptive driver-monitoring systems, as it can help identify early warning signs of drowsiness or distraction and allow for timely interventions. The clustering results  for the remaining subjects are provided in Table \ref{all_sub} in the Supplement.

\section{Discussion}\label{conclusion}
In this paper, we introduced FCPCA, a novel fuzzy clustering approach specifically designed for high-dimensional, variable-length  MTS. Unlike traditional distance-based fuzzy clustering \citep{egrioglu2013fuzzy,izakian2015fuzzy}, FCPCA measures reconstruction error, that is, how accurately each MTS is represented by a common subspace of each cluster. By shifting from distance minimization to a reconstruction-based criterion, FCPCA overcomes many pitfalls of centroid-based methods, as sensitivity to the curse of dimensionality.

Moreover, FCPCA employs a fuzzy membership framework that allows partial cluster assignments, capturing transitional or mixed states often overlooked by crisp clustering. Through iterative updates of both the membership matrix and the projection axes, FCPCA reveals subtle temporal relationships without forcing every series into a single cluster. This mechanism not only improves clustering accuracy but also provides a more nuanced representation of the underlying uncertainties in complex MTS data, outperforming traditional CPCA-based hard clustering methods.

Our experimental evaluations—spanning both synthetic simulations and real-world applications such as an EEG driver-drowsiness dataset—underscore the advantages of the proposed method. Notably, while numerous fuzzy and crisp clustering or classification methods have been applied to EEG-based driver-drowsiness detection \citep{gurudath2014drowsy, khushaba2010driver}, most focus solely on improving accuracy for well-defined states. By contrast, FCPCA reveals transitional states, thereby offering a more comprehensive understanding of driver vigilance.

Despite these promising results, our approach still has some limitations. While FCPCA is inherently less sensitive to outliers than centroid-based methods, it nevertheless relies on accurate covariance estimates, which can be affected by noisy or heavily corrupted data. Consequently, our future work includes extending FCPCA to a robust version that further mitigates this sensitivity. Looking ahead, FCPCA could also serve as a building block for other multivariate techniques, such as clustering-based forecasting.

The proposed FCPCA method is coded in R, and it is freely available on the following GitHub repository: \url{https://github.com/arbitraryma/FCPCA.git}.

\section*{Acknowledgments}
This research was supported by King Abdullah University of Science and Technology (KAUST). 


\section*{Conflict of interest statement}
The authors declare that they have no known competing financial
interests or personal relationships that could have appeared to
influence the work reported in this paper.

\section*{Data availability statement}
Data sharing is not applicable to this article, as the datasets used in our paper are already publicly available.

\pagebreak  
\newpage
\addcontentsline{toc}{chapter}{Bibliography}

\bibliographystyle{biom}
\bibliography{paper}

\newpage
        \section*{Supplement material}\label{supplement}

\begin{figure}[htbp]
    \centering
    \includegraphics[width=1\linewidth]{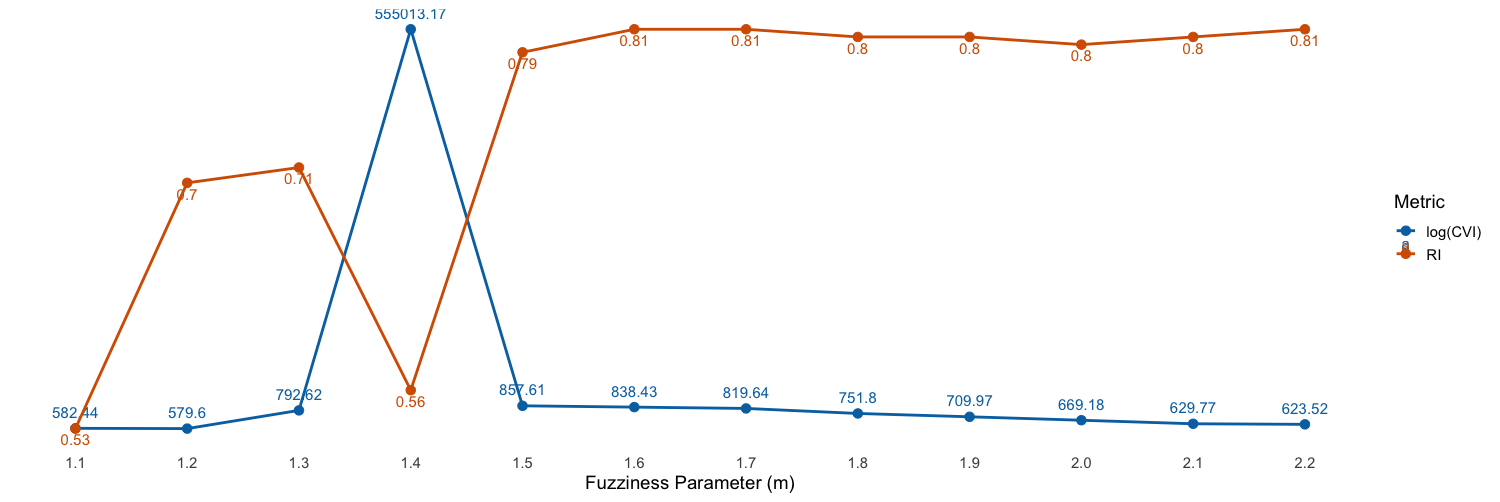}
    \caption{The selection of $m$ and comparison to the CVIs and RIs of subject 1}
    \label{subject1}
\end{figure}

\begin{figure}[htbp]
    \centering
    \includegraphics[width=1.0\linewidth]{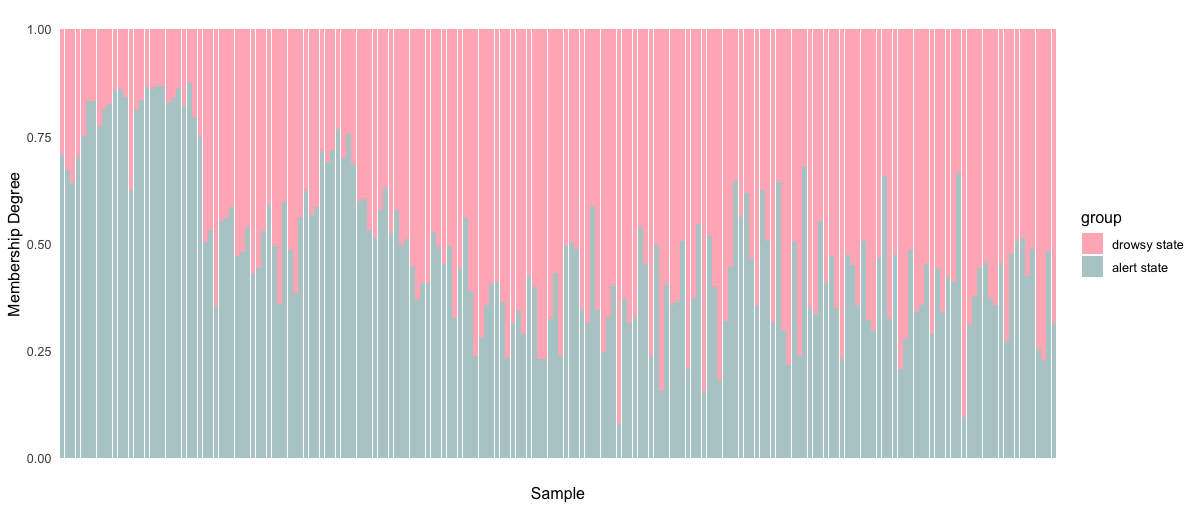}
    \caption{Membership matrix of the samples of subject 1.}
    \label{mem_1}
\end{figure}

\begin{figure}[htbp]
    \centering
    \includegraphics[width=1\linewidth]{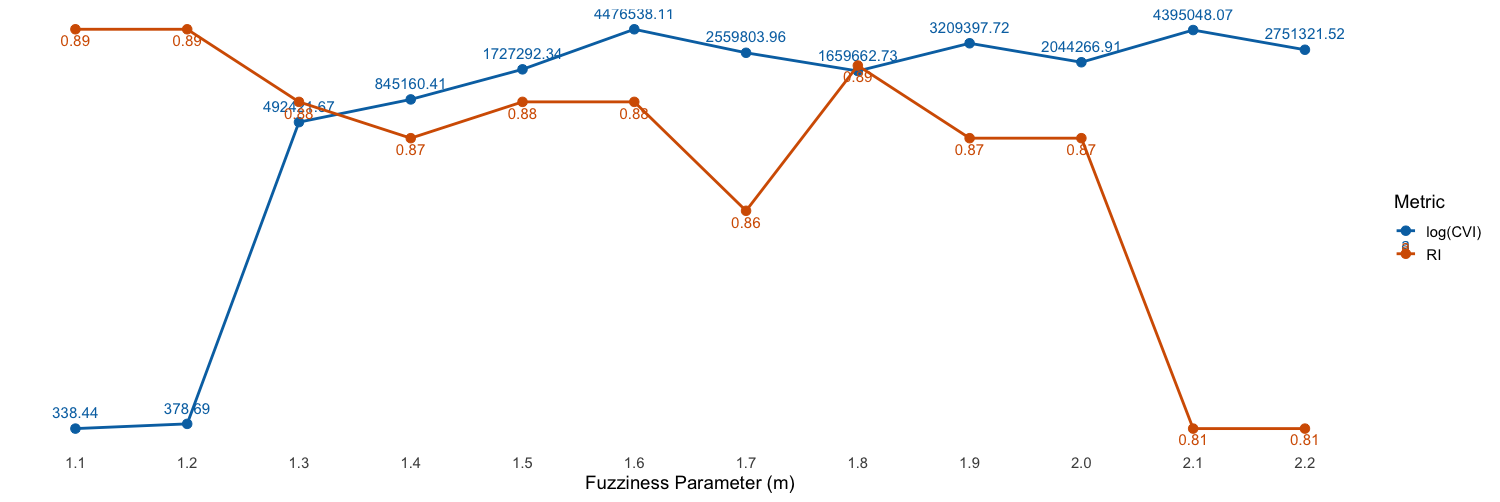}
    \caption{The selection of $m$ and comparison to the CVIs and RIs of subject 2.}
    \label{subject2}
\end{figure}

\begin{figure}[htbp]
    \centering
    \includegraphics[width=1.0\linewidth]{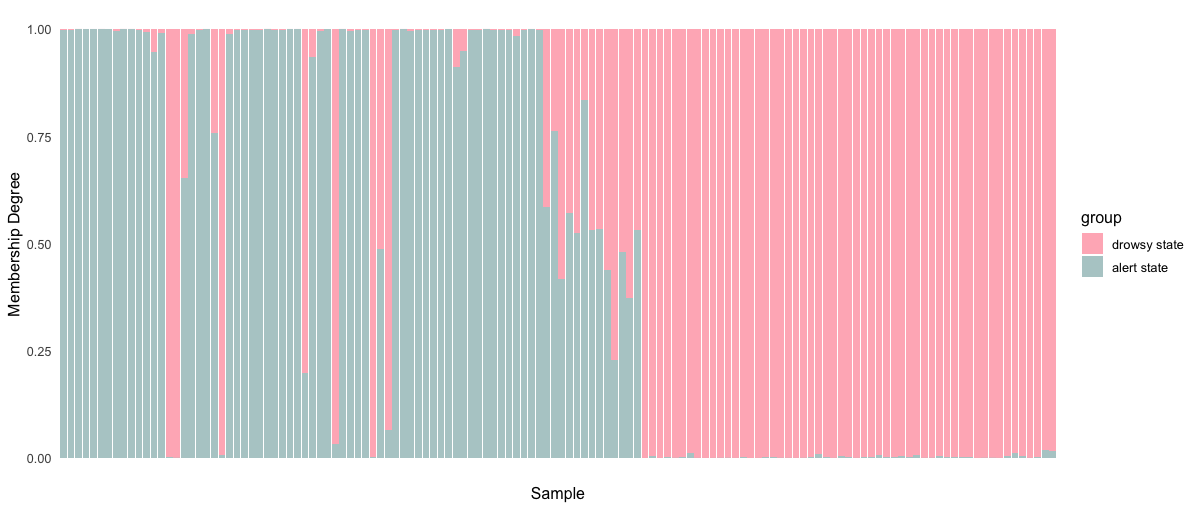}
    \caption{Membership matrix of the samples of subject 2.}
    \label{mem_2}
\end{figure}

\begin{figure}[htbp]
    \centering
    \includegraphics[width=1\linewidth]{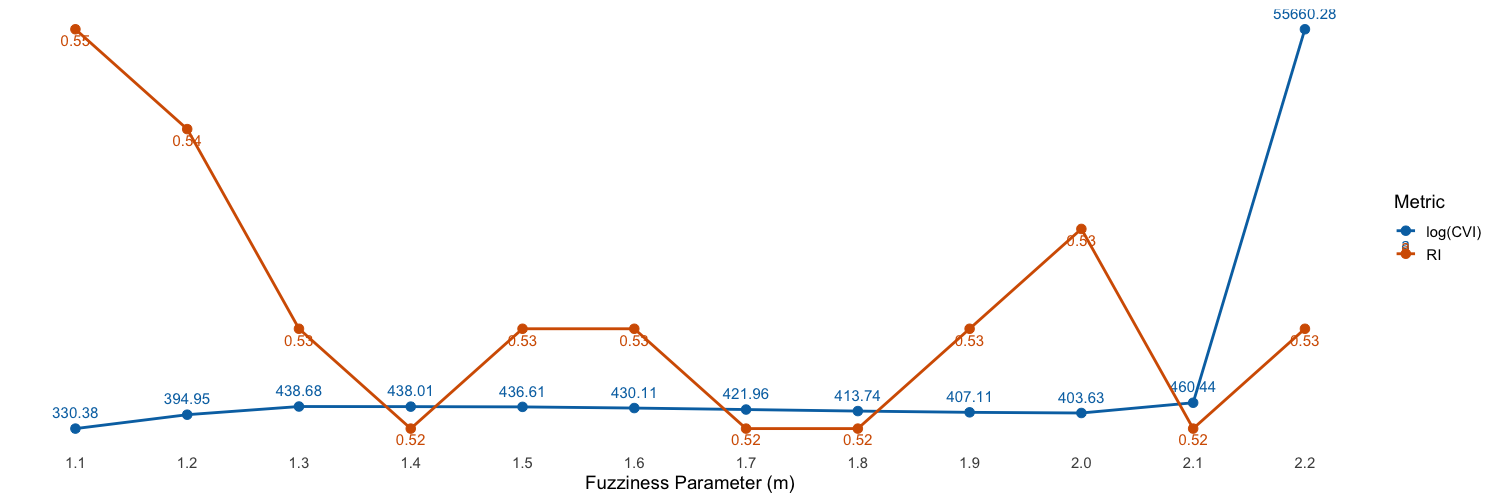}
    \caption{The selection of $m$ and comparison to the CVIs and RIs of subject 3.}
    \label{subject3}
\end{figure}

\begin{figure}[htbp]
    \centering
    \includegraphics[width=1.0\linewidth]{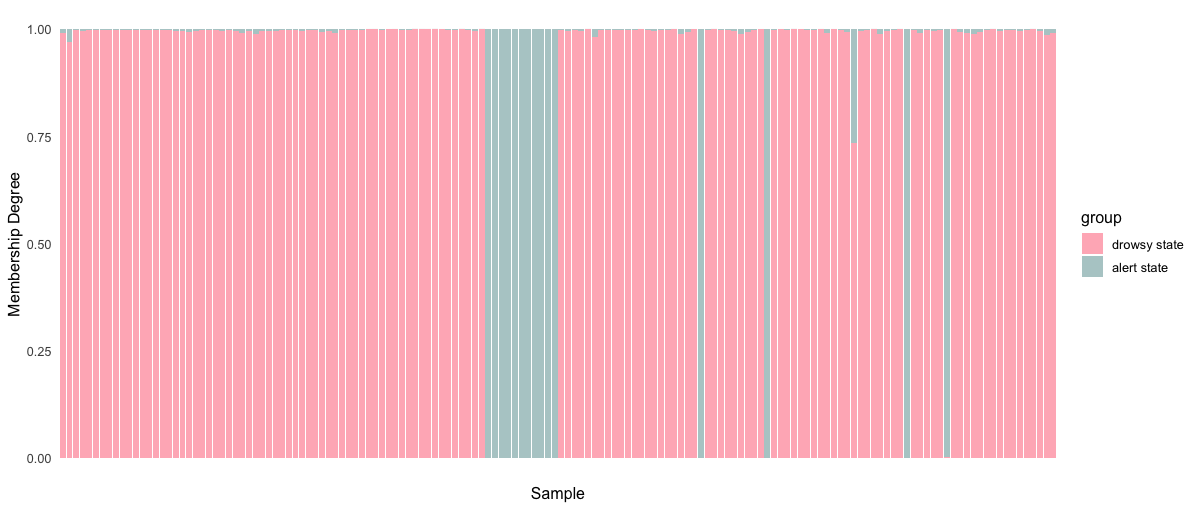}
    \caption{Membership matrix of the samples of subject 3.}
    \label{mem_3}
\end{figure}

\begin{figure}[htbp]
    \centering
    \includegraphics[width=1\linewidth]{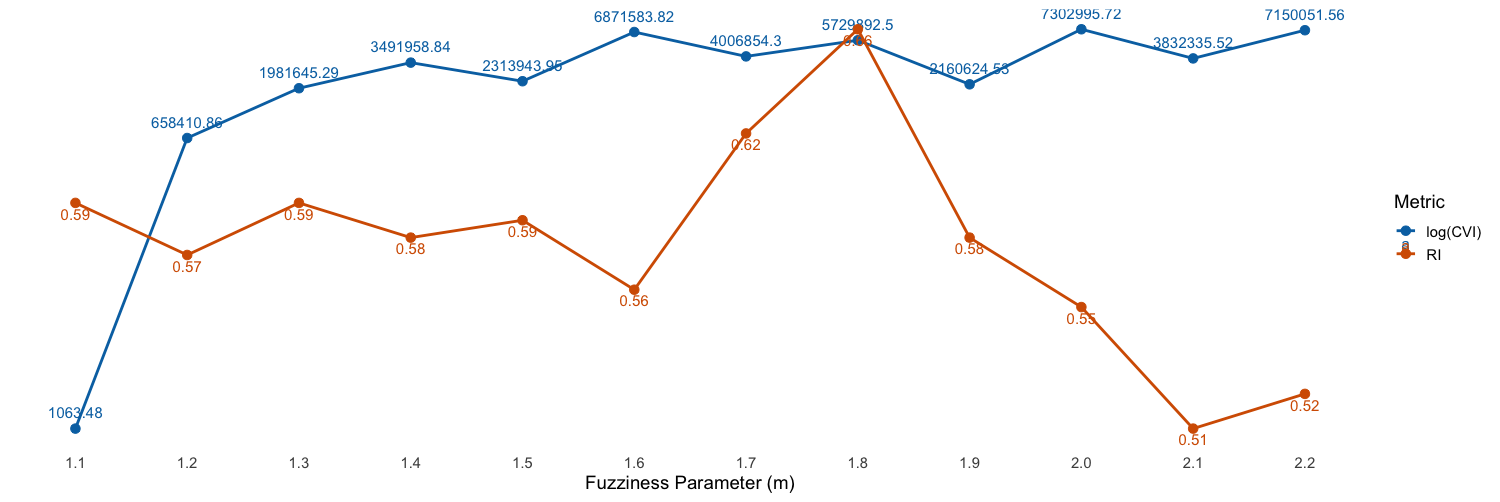}
    \caption{The selection of $m$ and comparison to the CVIs and RIs of subject 4.}
    \label{subject4}
\end{figure}

\begin{figure}[htbp]
    \centering
    \includegraphics[width=1.0\linewidth]{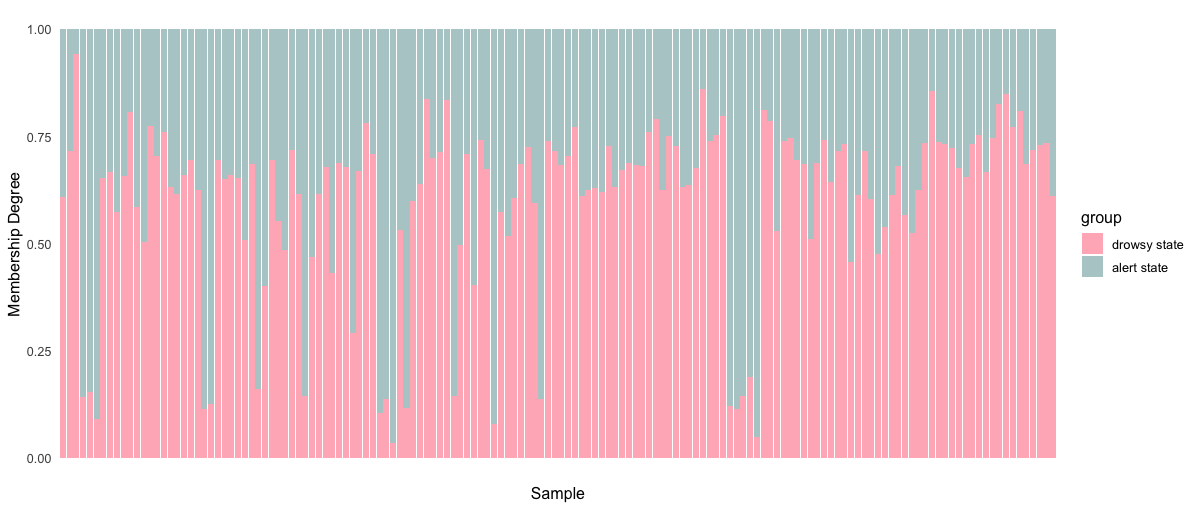}
    \caption{Membership matrix of the samples of subject 4.}
    \label{mem_4}
\end{figure}

\begin{figure}[htbp]
    \centering
    \includegraphics[width=1\linewidth]{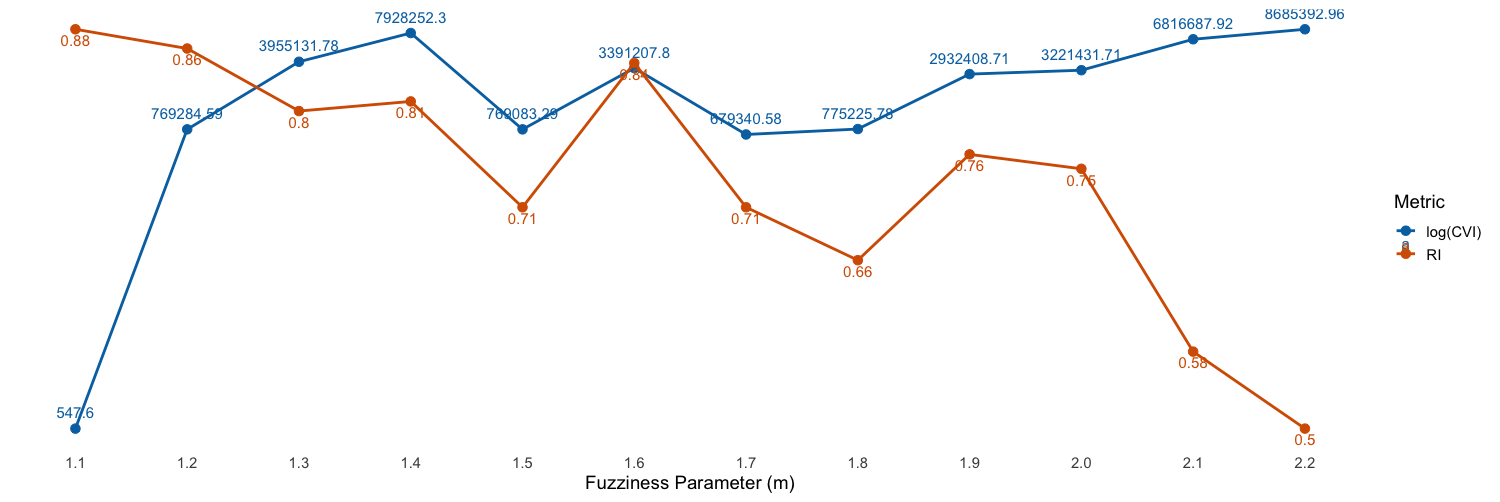}
    \caption{The selection of $m$ and comparison to the CVIs and RIs of subject 5.}
    \label{subject5}
\end{figure}

\begin{figure}[htbp]
    \centering
    \includegraphics[width=1.0\linewidth]{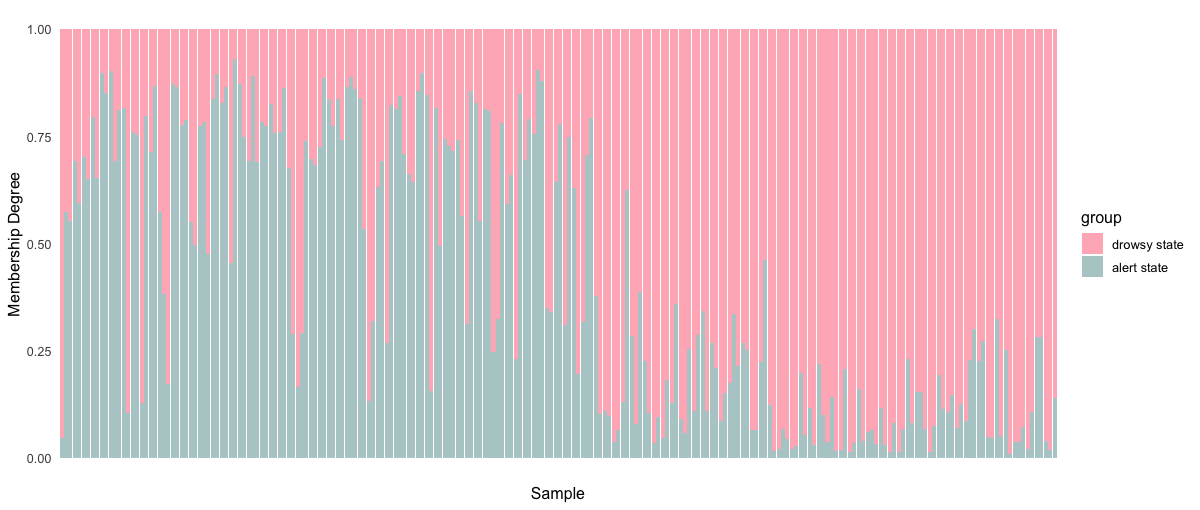}
    \caption{Membership matrix of the samples of subject 5.}
    \label{mem_5}
\end{figure}

\begin{figure}[htbp]
    \centering
    \includegraphics[width=1\linewidth]{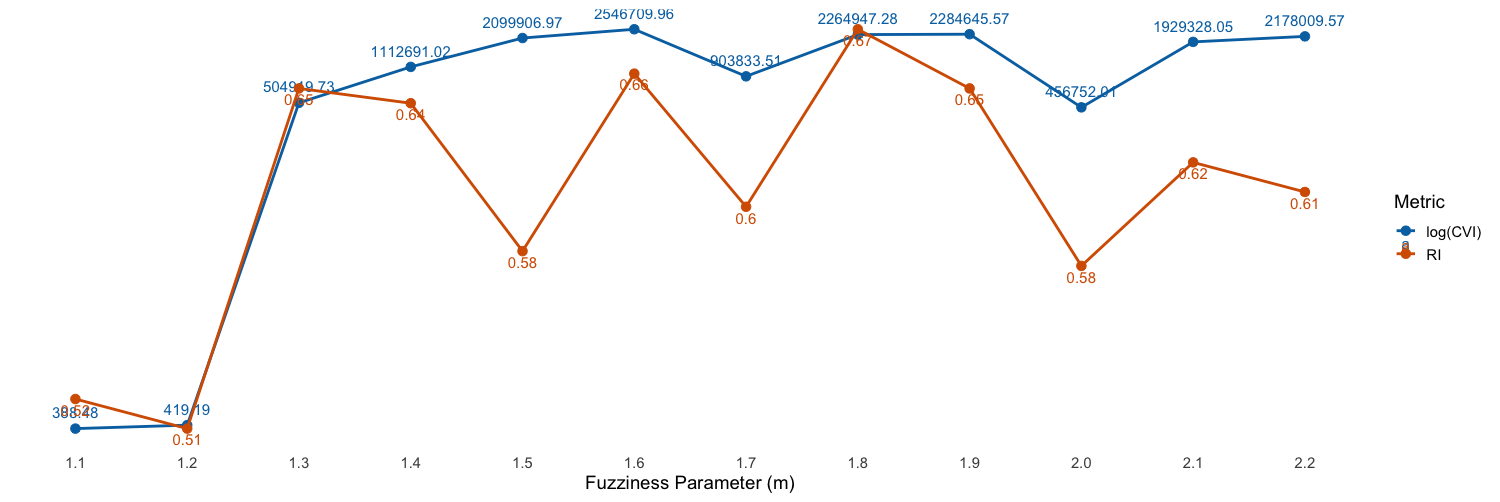}
    \caption{The selection of $m$ and comparison to the CVIs and RIs of subject 6.}
    \label{subject6}
\end{figure}

\begin{figure}[htbp]
    \centering
    \includegraphics[width=1.0\linewidth]{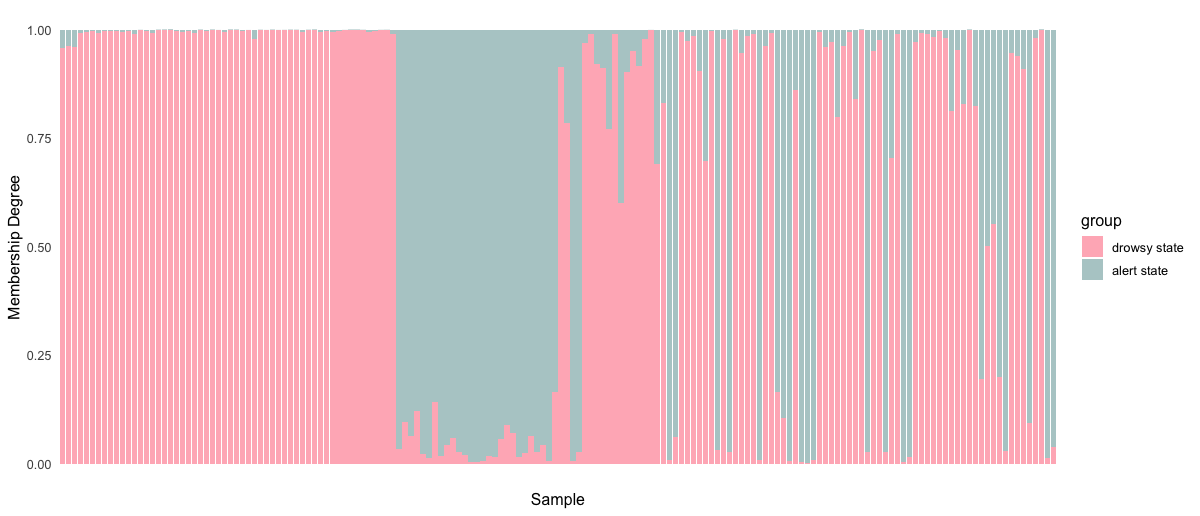}
    \caption{Membership matrix of the samples of subject 6.}
    \label{mem_6}
\end{figure}

\begin{figure}[htbp]
    \centering
    \includegraphics[width=1\linewidth]{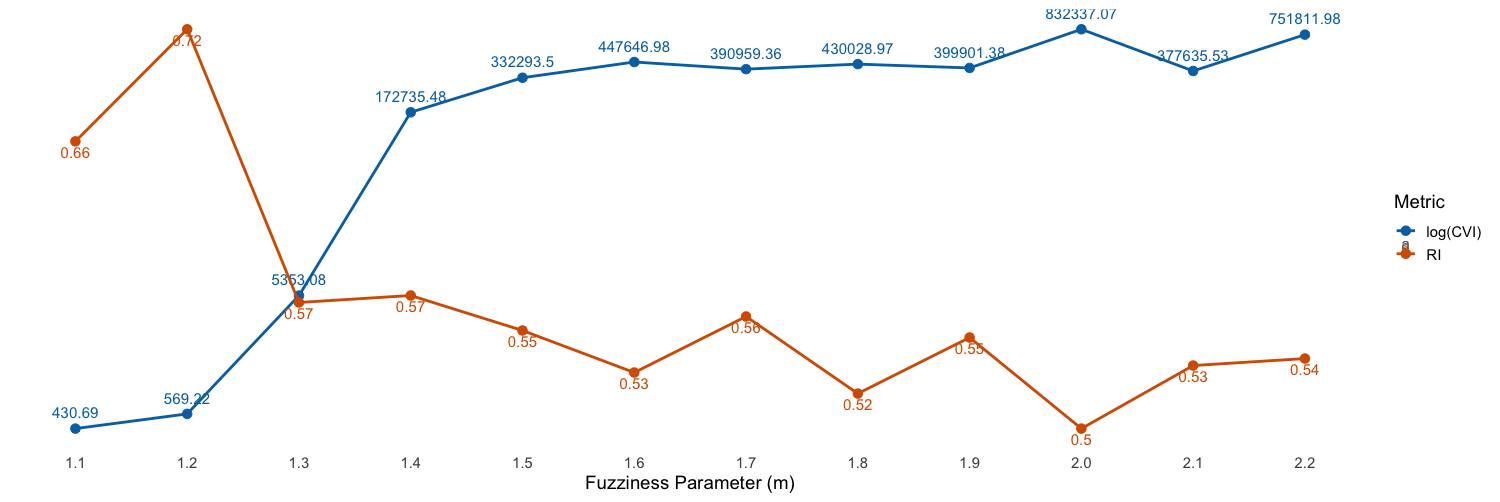}
    \caption{The selection of $m$ and comparison to the CVIs and RIs of subject 8.}
    \label{subject8}
\end{figure}

\begin{figure}[htbp]
    \centering
    \includegraphics[width=1\linewidth]{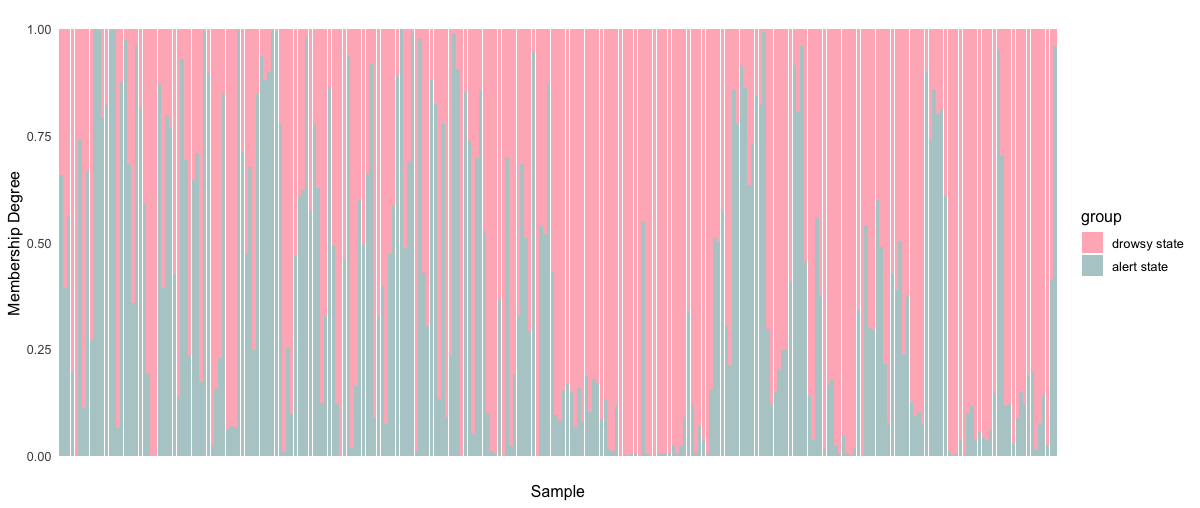}
    \caption{Membership matrix of the samples of subject 8.}
    \label{mem_8}
\end{figure}

\begin{figure}[htbp]
    \centering
    \includegraphics[width=1\linewidth]{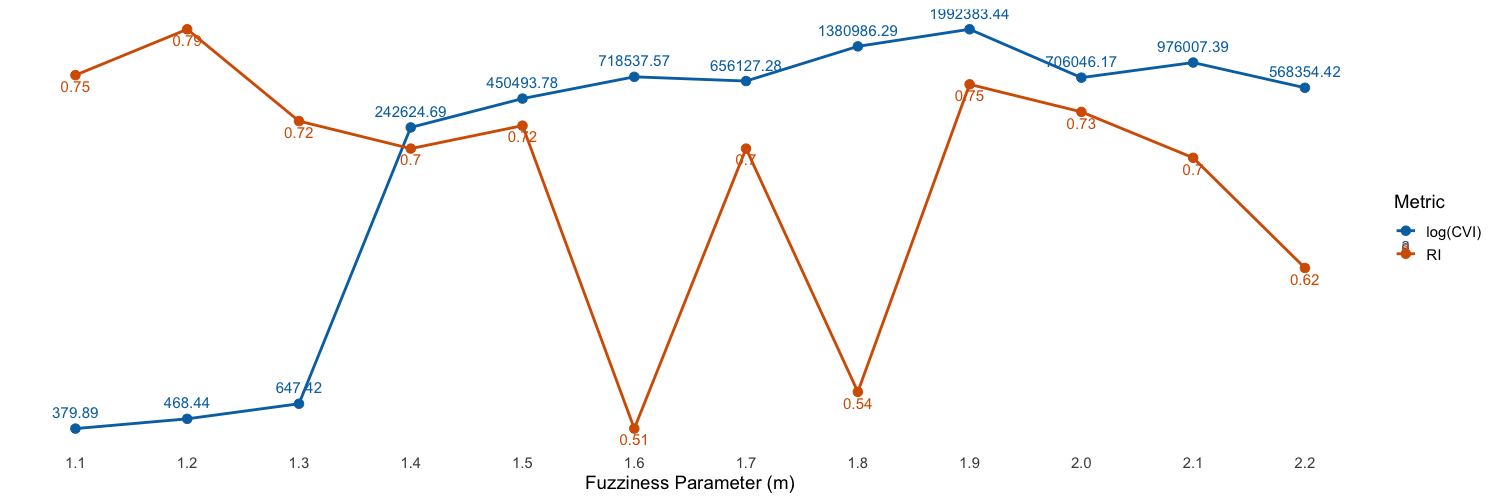}
    \caption{The selection of $m$ and comparison to the CVIs and RIs of subject 9.}
    \label{subject9}
\end{figure}

\begin{figure}[htbp]
    \centering
    \includegraphics[width=1.0\linewidth]{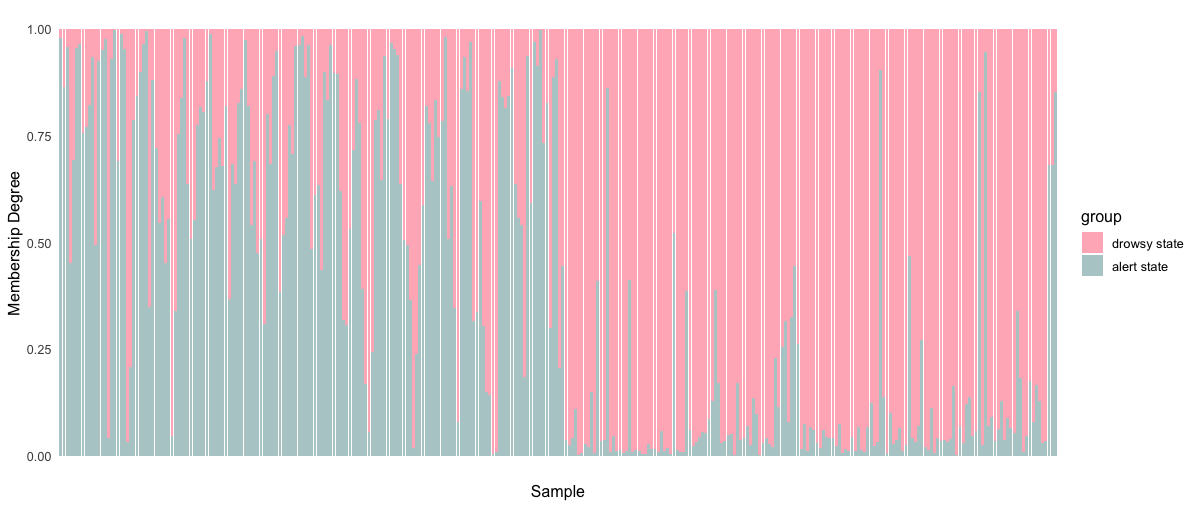}
    \caption{Membership matrix of the samples of subject 9.}
    \label{mem_9}
\end{figure}

\begin{figure}[htbp]
    \centering
    \includegraphics[width=1\linewidth]{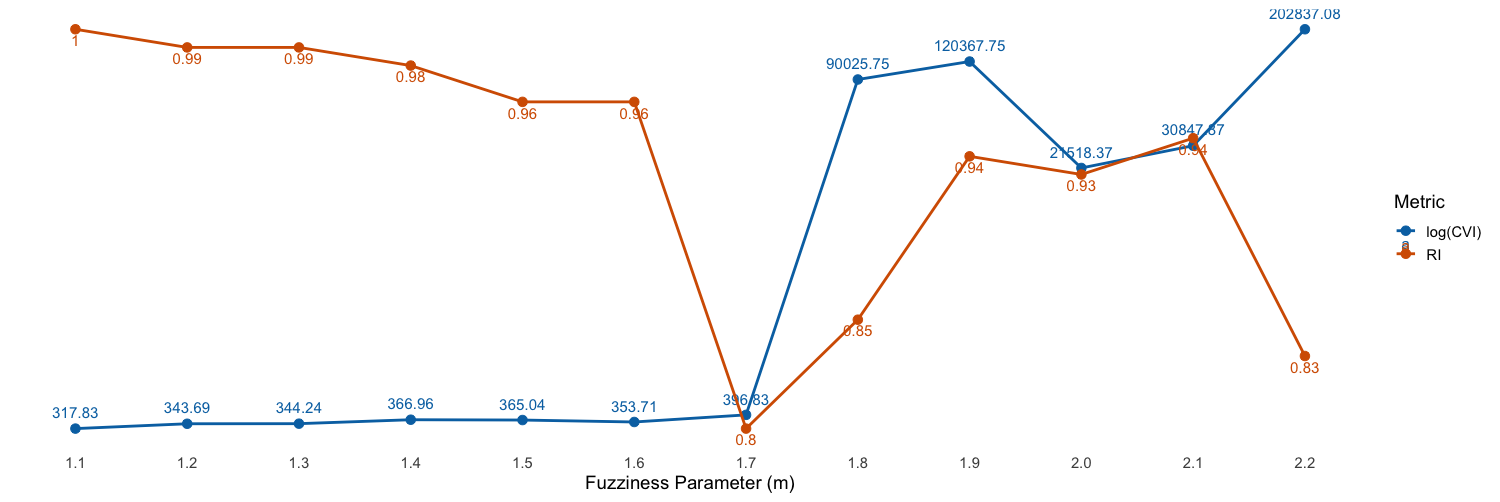}
    \caption{The selection of $m$ and comparison to the CVIs and RIs of subject 10.}
    \label{subject10}
\end{figure}

\begin{figure}[htbp]
    \centering
    \includegraphics[width=1.0\linewidth]{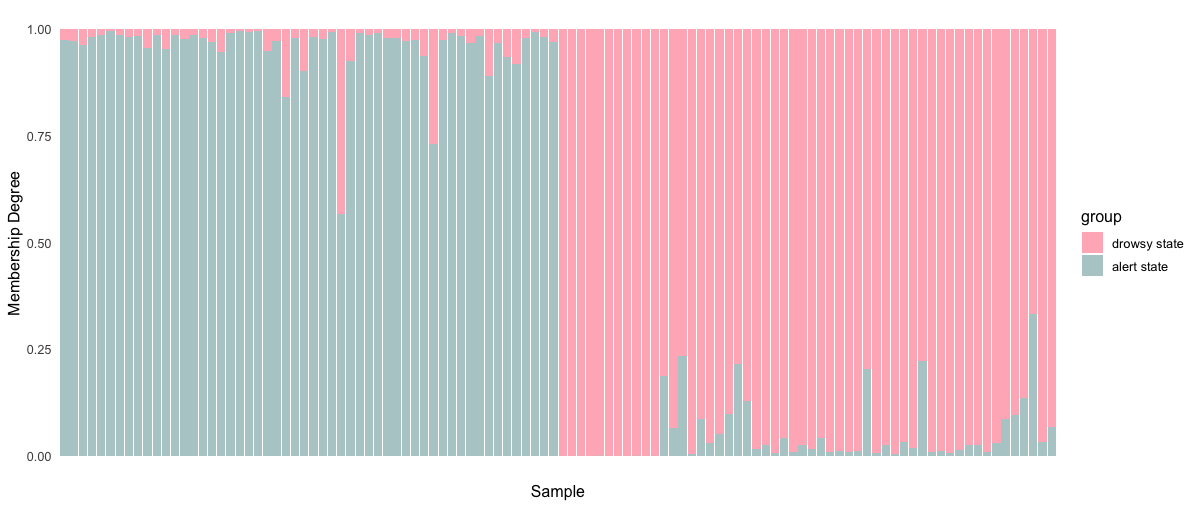}
    \caption{Membership matrix of the samples of subject 10.}
    \label{mem_10}
\end{figure}

In this section, we provide figures of the RIs according to the optimal selection of $m$ for the remaining 9 subjects (Table \ref{tables}). For each subject, we also provide the plots of the CVI-RI values and the membership matrix. These are shown from Figure \ref{subject1} to Figure \ref{mem_10}. For subject 10, our FCPCA achieves 100\% accuracy and delivers richer information over hard clustering methods, where one can observe the `transition' state on some samples. Note that our criterion for selecting the optimal fuzziness parameter $m$ works well.

\begin{table}[htbp]
    \centering
       \caption{RIs of each subject using FCPCA with the optimal selection of $m$}
       \label{tables}
 \begin{tabular}{c c c c }
 \hline
       Subject \#   & Sample size & Optimal $m$ & RI \\
       \hline
        1  & 188 & 1.2 & 0.70  \\
        2  & 132 & 1.1 & 0.90  \\
        3  & 150 & 1.1 & 0.55  \\
        4  & 148 & 1.1 & 0.59 \\
        5  & 224 & 1.1 & 0.88 \\
        6  & 166 & 1.1 & 0.52  \\
       7  & 102 & 1.1 & 0.58 \\
        8  & 264 & 1.1 & 0.66  \\
        9  & 314 & 1.1 & 0.75  \\
        10 & 108 & 1.1 & 1.00  \\
        11 & 226 & 1.1 & 0.93\\
        \hline
    \end{tabular}
 
    \label{all_sub}
\end{table}

\end{document}